\documentclass[12pt,a4paper]{article}
\pdfoutput=1

\usepackage{amsmath,amssymb,bbm} 
\usepackage{accents}
\usepackage{color}
\usepackage[bookmarks=true,hyperfigures=true]{hyperref}
\usepackage{graphicx}
\usepackage[nosort]{cite}
\usepackage[bulletsep]{collref}
\usepackage{tensor}

\usepackage[a4paper,text={173mm,216mm},centering]{geometry}

\allowdisplaybreaks[3]

\numberwithin{equation}{section}

\usepackage[font=small,labelfont=bf,width=0.85\textwidth]{caption}

\let\oldbfseries=\bfseries
\let\oldmdseries=\mdseries
\let\oldnormalfont=\normalfont
\renewcommand{\bfseries}{\oldbfseries\boldmath}
\renewcommand{\mdseries}{\oldmdseries\unboldmath}
\renewcommand{\normalfont}{\oldnormalfont\unboldmath}

\makeatletter
\newlength{\apb@width}
\newcommand{\autoparbox}[2][c]{\settowidth{\apb@width}{#2}\parbox[#1]{\apb@width}{#2}}

\makeatother

\newcommand{\nn}{\nonumber}

\DeclareMathOperator{\diag}{diag}                           
\newcommand{\be}{\begin{equation}}
\newcommand{\ee}{\end{equation}}
\newcommand{\ba}{\begin{eqnarray}}
\newcommand{\ea}{\end{eqnarray}}
\newcommand{\R}{\mathbb{R}}                                
\newcommand{\unit}{\mathbf{1}}                              

\ifx\genfrac\sdflkaj\else\fi
\newcommand{\sfrac}[2]{{\textstyle\frac{#1}{#2}}}
\newcommand{\half}{\sfrac{1}{2}}

\newcommand{\alg}[1]{\mathfrak{#1}}

\renewcommand{\a}{\alpha}
\newcommand{\da}{{\dot{\alpha}}}
\newcommand{\db}{{\dot{\beta}}}
\renewcommand{\b}{\beta}
\newcommand{\g}{\gamma}
\newcommand{\dg}{{\dot{\gamma}}}
\newcommand{\la}{\lambda}

\newcommand{\ft}[2]{{\textstyle\frac{#1}{#2}}}

\newcommand{\cO}{\mathcal{O}}

\def\l<{\langle}\def\r>{\rangle}

\newcommand{\eps}{\epsilon}

\newcommand{\vev}[1]{\langle#1\rangle}

\newcommand{\Tr}{\mathop{\mathrm{Tr}}}

\newcommand{\eqn}[1]{(\ref{#1})}
\newcommand{\btheta}{\bar\theta}
\newcommand{\dx}{{\dot x}}
\newcommand{\bpsi}{{\bar\psi}}

\newcommand*{\diff}{{\mathrm d}}

\begin{document}
\thispagestyle{empty}

\begingroup\raggedleft\footnotesize\ttfamily
HU-EP-13/42\\
NORDITA-2013-64\\
UUITP-10/13\\
\vspace{15mm}
\endgroup

\begin{center}
{\Large\bfseries Yangian Symmetry of smooth Wilson Loops in $\mathcal{N}=4$ 
super Yang-Mills Theory\par}%
\vspace{15mm}

\begingroup\scshape\large 
Dennis M\"uller, Hagen M\"unkler, Jan Plefka, \\ Jonas Pollok
\endgroup
\vspace{5mm}

\textit{Institut f\"ur Physik, Humboldt-Universit\"at zu Berlin, \phantom{$^\S$}\\
Newtonstra{\ss}e 15, D-12489 Berlin, Germany} \\[0.1cm]
\texttt{\small\{dmueller,muenkler,plefka,pollok\}@physik.hu-berlin.de\phantom{\ldots}} \\ \vspace{5mm}

\begingroup\scshape\large 
Konstantin Zarembo
\endgroup
\vspace{5mm}

\textit{Nordita, KTH Royal Institute of Technology and Stockholm University, \\
Roslagstullsbacken 23, SE-106 91 Stockholm, Sweden; \\
Department of Physics and Astronomy, \\ Uppsala University
SE-751 08 Uppsala, Sweden}\\[0.1cm]
\texttt{\small zarembo@nordita.org\phantom{\ldots}}
\vspace{8mm}


\textbf{Abstract}\vspace{5mm}\par
\begin{minipage}{14.7cm}
We show that appropriately supersymmetrized
smooth Maldacena-Wilson loop operators in $\mathcal{N}=4$ super Yang-Mills theory
are invariant under a Yangian symmetry $Y[\alg{psu}(2,2|4)]$
built upon the manifest superconformal symmetry algebra of the theory. 
The existence of this hidden symmetry 
is demonstrated at the
one-loop order in the weak coupling limit as well as at leading order in the
strong coupling limit employing the classical 
integrability of the dual $AdS_{5}\times S^{5}$ string description.
The hidden symmetry generators consist of a canonical non-local second order variational
derivative piece acting on the superpath, along with a novel local path dependent contribution. 
We match the functional form of these Yangian symmetry generators at weak and strong
coupling and find evidence for an interpolating function.
Our findings represent the smooth counterpart to the Yangian invariance of 
scattering superamplitudes dual to light-like polygonal super Wilson loops in the 
 $\mathcal{N}=4$ super Yang-Mills theory.

\end{minipage}\par
\end{center}
\newpage


\setcounter{tocdepth}{2}
\hrule height 0.75pt
\tableofcontents
\vspace{0.8cm}
\hrule height 0.75pt
\vspace{1cm}

\setcounter{tocdepth}{2}

\section{Introduction and summary}

The $\mathcal{N}=4$ supersymmetric Yang-Mills (SYM) theory is a distinguished four dimensional
gauge theory that has been intensively studied in recent years. It is a finite, maximally supersymmetric and quantum superconformal four dimensional gauge theory which may be
understood as an idealized version of QCD. 
In the planar limit the model with $SU(N)$ gauge group exhibits integrability, which manifests 
itself through an infinite dimensional extension of the superconformal symmetry algebra of Yangian
type. 
Not being a symmetry of the action integrability appears in 
gauge invariant observables of the theory with a non-trivial dependence on the 't Hooft coupling $\lambda$. Of course $\mathcal{N}=4$ SYM theory is also the gauge theory with
the best established string theory dual description in terms of the 
$AdS_{5}\times S^{5}$ superstring. 
The string dynamics,  described by a two-dimensional quantum field theory on the  worldsheet, also appears to be integrable. 

So far integrable structures have been detected in  $\mathcal{N}=4$ SYM
 for two- and three-point functions 
of local gauge invariant operators. Our understanding for the case of two-point 
functions which yield the
local operator's scaling dimensions is the most mature. Exact results are obtained through
a reformulation of the problem in terms of a dynamic super-spin chain, associated
Bethe ans\"atze and their generalizations, see \cite{Beisert:2010jr} for a comprehensive overview. 
This language was also successfully used recently for the study of three-point
functions \cite{Escobedo:2010xs,Escobedo:2011xw,Gromov:2011jh,Foda:2011rr,Kazakov:2012ar,
Foda:2013nua}. A further sector is that of scattering amplitudes in the gauge theory which
are invariant under superconformal transformations, see \cite{Drummond:2010km} for an introduction. 
Here the discovery of a hidden dual superconformal symmetry \cite{Drummond:2008vq}, which combines with the conventional superconformal symmetry 
into a Yangian symmetry algebra \cite{Drummond:2009fd} again points to an underlying integrability. The Yangian invariance
of tree-level super-amplitudes in the theory was argued to extend to the loop-level
integrands in \cite{ArkaniHamed:2010kv}. The one-loop amplitudes enjoy a deformed symmetry
\cite{Beisert:2010gn,Sever:2009aa}. In $\mathcal{N}=4$ SYM theory scattering amplitudes are
dual to supersymmetrically extended Wilson loops with light-like polygonal boundaries 
\cite{Alday:2007hr,Brandhuber:2007yx,Drummond:2007cf,CaronHuot:2010ek,Mason:2010yk,Belitsky:2012nu,CaronHuot:2011ky,Beisert:2012xx}. A recent constructive
application of integrability to the space-time S-matrix uses a decomposition of the dual
Wilson loop into pentagon blocks and yields 
non-perturbative results in the form of an OPE-like expansion \cite{Basso:2013vsa,Basso:2013aha}. 
All these developments point towards  rich integrable structure behind amplitudes/null-polygon Wilson loops, which is not completely uncovered yet, partly due to breakdown or deformation of the superconformal and Yangian symmetries by
the infrared (amplitudes) or ultraviolet (light-like Wilson loops) divergencies.

In this paper we turn to a prominent further class of observables in $\mathcal{N}=4$ SYM 
and its string-dual  being almost
as old as the AdS/CFT correspondence proposal \cite{Maldacena:1997re} itself : The Maldacena-Wilson
loop operators \cite{Maldacena:1998im,Rey:1998ik}. 
Here the loop-path variable couples next to the gauge field 
also to the adjoint scalars in the form \cite{Maldacena:1998im}
\begin{align}
W(C)= \frac{1}{N} \mathrm{Tr} \: \mathcal{P} \exp{ \left( i \oint_C \mathrm{d} \tau  \left( A_{\mu}(x) \dot{x}^{\mu} + \Phi_i (x)|\dot{x}| n^i \right) \right)}\qquad\text{with } ~(n^{i})^2=1 \, . \label{eins}
\end{align} 
The Maldacenca-Wilson loop operators are locally $1/2$ BPS symmetric, their
expectation values are finite for smooth loops 
and are invariant under conformal transformations.
The expectation value at strong coupling follows from the regularized
minimal surface of an open string in anti-de-Sitter space ending on the four dimensional 
boundary on the curve $C$ of the Wilson loop. 
As a direct consequence of integrability of the differential equations that determine the shape of the minimal surface, the minimal area satisfies a number of Ward identities of Yangian type which can be derived from the Hamilton-Jacobi formalism \cite{Polyakov:2000ti,Polyakov:2000jg} following an unpublished idea of Polyakov  \cite{Polyakov-unpublished}.
For the expectation value of the circular Maldacena-Wilson loop exact results to all orders in $\lambda$ and
$1/N$ are available \cite{Erickson:2000af,Drukker:2000rr}. Given these properties it is natural to
ask whether integrability in the sense of a hidden Yangian symmetry exists for 
smooth Maldacena-Wilson loops both at weak and at strong coupling\footnote{This idea was jointly developed with N.~Drukker,
as well as independently with A.~Sever and P.~Vieira.}. 
Our goal is to investigate possible Yangian symmetries of Wilson loops in detail.

As we will discuss the conformal symmetry for the Maldacena-Wilson loop is represented through 
functional derivative operators acting on the space of paths $x^{\mu}(\tau)$ e.g.~for the special
conformal transformations
\begin{align*}
\int ds \: k^{\mu}(s)\, \vev{W(C)}= 0 \, , \qquad \text{with} \quad
k^{\mu}(s)= x^2(s) \dfrac{\delta}{\delta x^{\mu}(s)}- 2 x_{\mu}(s) x^{\nu}(s) \dfrac{\delta}{\delta x^{\nu}(s)} \, ,
\end{align*}
and similarly for the dilatations $d(s)$ as well as Poincar\'e transformations.
In order to establish
the Yangian symmetry it turns out to be necessary to consider the supersymmetrization
of the Maldacena-Wilson loop operator \eqn{eins} describing a path in an \emph{non}-chiral superspace
$\{ x^{\mu}(\tau), \theta^{A}_{\alpha}(\tau), \bar{\theta}_{A\, \dot\alpha}(\tau) \}$
with $\a,\da=1,2$ and $A=1,2,3,4$.
We establish this object up to second-order in anti-commuting path variables and
show its superconformal invariance at leading order in perturbation theory. One can think
of it as a smooth version of the light-like polygonal non-chiral super-Wilson loops of 
\cite{CaronHuot:2011ky,Beisert:2012xx} although we have not yet detailed the
precise relation.
At weak and strong coupling  we show that a natural definition of the
level-one generators of the Yangian algebra of $\alg{psu}(2,2|4)$ indeed annihilate
the constructed super Maldacena-Wilson loops $\vev{\mathcal{W}(C)}$. Concretely for
 the level-one momentum
generator $P^{(1)}_{\mu}$ we show
\begin{align}
\Biggl \{
\int_{s_{1}<s_{2}} \! \! \! \! \! \! \! \! \mathrm{d} s_1   \mathrm{d} s_2 \:
\Bigl(\, & d(s_{1})\, \eta^{\mu\nu} - m^{\mu\nu}(s_{1})\, \Bigr)\, p_{\nu}(s_{2})
- \frac{i}{4}  \, {\bar q}^{A\, \da}(s_1)  \, \bar{\sigma}^{\mu}_{\a\da}\,  q_A^{\a}(s_2)
 -  
\left( s_1 \leftrightarrow s_2 \right) \nn\\
& + f(\lambda)\,\int ds \left ( \frac{\ddot{x}^{2}}{\dot{x}^{4}} - \frac{(\dot{x}\cdot\ddot{x})^{2}}{\dot{x}^{6}} 
\, \right ) \dot{x}^{\mu} \Biggr\} \,  \vev{\mathcal{W}(C)} =0  
\label{zwei}
\end{align}
at leading order in the weak or strong coupling expansion. Assuming that there are no fermionic
corrections at the leading order of strong coupling expansion our results indicate that the
function $f(\lambda)$ has limiting behavior 
$$
f(\lambda\ll 1)= \frac{7 \, \lambda}{96\, \pi^{2}}\, , \quad \text{and}\quad
f(\lambda\gg 1)= \frac{\lambda}{4\, \pi^{2}}\, ,
$$
at weak and strong coupling. It would be interesting to understand the form of $f(\lambda)$ beyond
these leading orders. Hence in this paper we provide good evidence for the existence of an infinite
dimensional hidden symmetry of the super Maldacena-Wilson loops
\be
J_{a}^{(n)}\, \vev{\mathcal{W}(C)} =0\, , \qquad J_{a}^{(n)}\in Y[\alg{psu}(2,2|4)]\,, \quad
 n\in\mathbb{N}\,  .
\label{drei}
\ee

We note that this uncovered hidden symmetry \eqn{drei} has strong similarities to 
a more than 30 year old speculation of Polyakov \cite{Polyakov:1979gp,Polyakov:1980ca} on the existence of a hidden symmetry for Wilson loops in pure Yang-Mills theory  related to the integrability
of the non-linear sigma model. In a sense \eqn{drei} is a realization of this
for $\mathcal{N}=4$ SYM. Parallel to these works \cite{Polyakov:1979gp,Polyakov:1980ca}
loop equations were proposed by  Makeenko and Migdal \cite{Makeenko:1979pb,Makeenko:1980vm}
in a related attempt to reformulate QCD as the dynamics of Wilson loops. 
Indeed generalizations of the
loop equations to super Maldacena-Wilson loops in the AdS/CFT context were studied before
in \cite{Drukker:1999zq,Drukker:1999gy,Polyakov:2000ti}. The Yangian
symmetry generators \eqn{drei} are somewhat different from the loop Laplacian   appearing in the loop equations. First of all,  the Yangian  generators are non-local, unlike the Laplacian. They are also honest
second-order variational operators and, in contradistinction to the loop Laplacian, do not satisfy the Leibniz rule. In the Makeenko-Migdal equations self-intersecting loops are very important. On the other hand Yangian and conformal symmetries in general get broken by cusps or self-intersections, and in this paper we only consider smooth, non-intersecting loops.

\section{Maldacena-Wilson Loop}

The Maldacena-Wilson Loop operator in $\mathcal{N}=4$ Super-Yang Mills Theory in Lorentzian
signature $\eta^{\mu \nu}=$diag$(+,-,-,-)$ is given by 
(\ref{eins}),
where $x^\mu(s):$ $[a,b]\rightarrow \mathbb{R}^{1,3}$ parametrizes the integration contour $C$,
$\mathcal{P}$ denotes path-ordering and $n^i$ is a constant unit $6$-vector which specifies a point on $S^5$. Note also that we define the modulus as $|\dot{x}| := \sqrt{\dot{x}^{2}}$, hence for space-like velocity  $|\dot{x}|$ is imaginary, i.e.~$|\dot{x}|=i\,  \lVert \dot{x} \rVert $.
At leading order in perturbation theory one easily derives the correlation function
\begin{align}
\left\langle   {W}(C)  \right\rangle = 1 - \frac{\lambda}{16 \pi^2 }  \int & \diff \tau_1    \diff \tau_2  \: \frac{\dot{x}_1\cdot \dot{x}_2 - |\dot{x}_1| |\dot{x}_2|  }{(x_1-x_2)^2} + \ldots \mbox{ ,} \label{wilson1}
\end{align} 
where we have abbreviated $x_i := x(\tau_i)$, 
see appendix \ref{app:Conventions} and \ref{app:Propagators} for our conventions and propagators. 
In our discussion of the symmetries of the Maldacena-Wilson loop we will initially focus on the one-loop contribution for which we introduce the notation
\begin{align}
\left\langle W(C) \right\rangle_{(1)} = - \frac{\lambda}{16 \pi^2 } \int \diff \tau_1 \diff \tau_2 \: I_{1 2}   \quad \mbox{ where} \quad I_{1 2} := \frac{\dot{x}_1  \dot{x}_2 - |\dot{x}_1| |\dot{x}_2|}{(x_1-x_2)^2} \, .\label{loop}
\end{align}

\subsection{Conformal invariance at one-loop}
\label{sect:2.1}

Before our discussion of possible hidden symmetries of the Maldacena-Wilson Loop we review its invariance under conformal transformations. 
We introduce the following natural functional derivative representation of the conformal algebra acting on the path $x^{\mu}(s)$ of the loop contour $C$ 
\begin{align}
P_{\mu}& =  \int ds\,  p_{\mu}(s) =  \int \diff s \: \dfrac{\delta}{\delta x^{\mu}(s)} \nn\\
M_{\mu \nu} & =  \int ds\,  m_{\mu\nu}(s)=  \int \diff s \left( x_{\mu}(s) \dfrac{\delta}{\delta x^{\nu}(s)} - x_{\nu}(s) \dfrac{\delta}{\delta x^{\mu}(s)} \right) \nn\\
D& =  \int ds \, d(s)=  \int \diff s \: x^{\mu}(s) \dfrac{\delta}{\delta x^{\mu}(s)} \nn\\
K_{\mu} & =  \int ds\,  k_{\mu}(s) =  \int \diff s \: \left( x^2(s) \dfrac{\delta}{\delta x^{\mu}(s)}- 2 x_{\mu}(s) x^{\nu}(s) \dfrac{\delta}{\delta x^{\nu}(s)} \right)  \, , 
\label{densities1}
\end{align}
introducing the momentum and angular momentum densities $p_{\mu}(s)$ and $m_{\mu\nu}(s)$
as well as the dilatation $d(s)$ and special conformal density $k_{\mu}(s)$. These generators satisfy the commutation relations (\ref{conf_algebra}).

Let us now prove the conformal invariance of the one-loop expectation value $\left\langle W(C) \right\rangle_{(1)}$. For this we note the functional derivatives
\begin{align}
\dfrac{\delta x_{\nu}(\tau)}{\delta x^{\mu}(s)}  = \eta_{\mu \nu} \delta(\tau -s)\, ,
\quad
\dfrac{\delta \dot{x}_{\mu}(\tau) }{\delta x^{\nu}(s)} = \eta_{\mu \nu} \, \partial _{\tau} \, \delta \left(\tau - s \right) \, ,\quad
\dfrac{\delta |\dot{x}(\tau)| }{\delta x^{\nu}(s)} = \frac{\dot{x}_{\nu}(\tau)}{|\dot{x}(\tau)|}\, \partial _{\tau} \, \delta \left(\tau - s \right) \, . \label{rule1}
\end{align} 
It comes as no surprise that already the integrand $I_{12}$ of $\left\langle W(C) \right\rangle_{(1)}$ is 
translation invariant by acting on it with $P_{\mu}$
\begin{align*}
P_{\mu}\, I_{1 2} & =  \int \diff s  \, \dfrac{\delta}{\delta x^{\mu}(s)} \frac{\dot{x}_1  \dot{x}_2 - |\dot{x}_1| |\dot{x}_2|}{(x_1-x_2)^2} \\
& =  \frac{1}{\left(x_1-x_2\right)^2} \bigg[ \bigg( \, \dot{x}_{\mu}(\tau_1) 
- \frac{|\dot{x}(\tau_1)|}{|\dot{x}(\tau_2)|}\dot{x}_{\mu}(\tau_2)\, \bigg)
 \int \diff s  \, \partial_{\tau_2} \delta (\tau_2 - s) + (1\leftrightarrow 2) \bigg ]
\\
& \quad - 2 \, \frac{\dot{x}_1  \dot{x}_2 - |\dot{x}_1| |\dot{x}_2|}{(x_1-x_2)^4} \, ( x_{1\,\mu} -  x_{2\,\mu} )  \int \diff s  \,( \delta(\tau_1 - s) - \delta(\tau_2 - s))
\end{align*}  
By virtue of
\be
\partial _{\tau_i} \int \diff s  \: \delta (\tau_i - s) = 0
\qquad \text{and} \qquad \int \diff s \: ( \delta(\tau_1 - s) - \delta(\tau_2 - s)) = 0
\ee
we indeed find invariance at the integrand level
\begin{align*}
P_{\mu}\,  I_{1 2}  = 0 \, .
\end{align*}  
A similar computation reveals that $ M_{\mu \nu}\, I_{1 2} = 0$ and the scale invariance
of $I_{12}$
is manifest. For the generators $ K_{\mu}$ of special conformal transformations the computation is  a bit more involved. We first calculate the action of $K_{\mu}$ on the integrand $I_{1 2}$
writing $x_{12}=x_{1}-x_{2}$
\begin{align*}
& K_{\mu}\, I_{1 2}  =  \int \diff s \, \left( x^2 (s) \delta^{\nu} _{\mu} - 2 x_{\mu}(s) x^{\nu}(s) \right) \dfrac{\delta}{\delta x^{\nu}(s)} \frac{\dot{x}_1  \dot{x}_2 - |\dot{x}_1| |\dot{x}_2|}{(x_1-x_2)^2} \\ 
& =- 2 \left( \dot{x}_{1,\mu} \frac{\dot{x}_2 x_{12}}{x_{12}^2} - \dot{x}_{2,\mu} \frac{\dot{x}_1 x_{12}}{x_{12}^2} \right) 
 =    \dot{x}_{1,\mu} \, \partial_2 \ln \left(-x_{12}^2\right) + \dot{x}_{2,\mu} \,  \partial_1 \ln \left(-x_{12}^2 \right)  \, ,
\end{align*} 
which is a total derivative in each term.
Therefore the integrated expression for a closed loop is invariant and we have
\begin{align*}
K_{\mu}\, \left\langle W(C) \right\rangle_{(1)} = 0 \mbox{ ,} 
\end{align*} 
as claimed.
In fact the conformal invariance of $\vev{ W(C)}$ may be shown via Ward identities
beyond perturbation theory.

\subsection{Towards a hidden Yangian symmetry}
\label{sect:Yangiansym}

Inspired by the Yangian symmetry discovered for scattering amplitudes 
in $\mathcal{N} = 4$ SYM \cite{Drummond:2009fd} which are dual to
light-like supersymmetric Wilson loops \cite{Alday:2007hr,Brandhuber:2007yx,Drummond:2007cf,CaronHuot:2010ek,Mason:2010yk,Belitsky:2012nu,Beisert:2012xx}
it is natural to search for a parallel structure for the
Maldacena-Wilson loop. 

Yangian symmetries traditionally appear in 2d integrable field theories 
(see \cite{MacKay:2004tc} for a review)
and in fact our
construction at strong coupling to be discussed in section \ref{sect:sc} follows this.  
The Yangian algebra $Y(\alg{g})$ of a simple Lie algebra $\alg{g}$ was
introduced by Drinfeld \cite{Drinfeld:1985rx,Drinfeld:1986in}. It is a deformation of the
loop-algebra spanned by the generators $J_{a}^{(n)}$  with grading $n\in\mathbb{N}$. 
One demands the level-zero and level-one commutation relations
\be
\label{comrels}
[ J_{a}^{(0)}, J_{b}^{(0)}\} = f_{ab}^{c}\, J_{c}^{(0)} \, ,\qquad
[ J_{a}^{(0)}, J_{b}^{(1)}\} = f_{ab}^{c}\, J_{c}^{(1)} \, .
\ee
where we use mixed brackets $[.,.\}$ to denote the graded commutator. 
In fact the first two level generators $J_{a}^{(0)}$ and $J_{a}^{(1)}$ span all of
$Y(\alg{g})$.
In addition there
is a set of representation dependent Serre relations, a generalized Jacobi-like identity.
The higher level generators follow from commutators of the level-one generators.
The distinguishing feature of the Yangian is a non-trivial co-product for the level-one
generators
\be
\label{coproducts}
\Delta(J^{(0)}_{a}) = J^{(0)}_{a} \otimes \unit + \unit \otimes J^{(0)}_{a} \, , \qquad
\Delta(J^{(1)}_{a}) = J^{(1)}_{a} \otimes \unit + \unit \otimes J^{(1)}_{a} 
+ f_{a}{}^{bc}\, J^{(0)}_{b} \otimes J^{(0)}_{c}\, . \qquad
\ee
Note that in the last term quadratic in $J^{(0)}_{a}$ the structure constant with `inverted'
indices appears. Indices are raised and lowered with the group metric $\half \Tr(J^{(0)}_{R\, a}\,
J^{(0)}_{R\,b})$ with $J^{(0)}_{R, a}$  in the defining representation of $\alg{g}$.

This is known as Drinfeld's first realization of the Yangian.
Importantly however, the Yangian is closely related to the Yang-Baxter equation which is 
central to the quantum inverse scattering method of integrability. 
$Y(\alg{g})$ may be also given a Hopf algebra structure by introducing a co-unit and an antipode,
but that will be of no relevance here.

An integrable 2d field theory realizes these structures in physics. The above co-products
translate  
 \be
J_{a}^{(0)}=\int ds \, j_{a}^{(0)}(s)\, , \qquad
J_{a}^{(1)}=\int ds\,  j_{a}^{(1)}(s) + f_{a}{}^{bc}\int_{s_{1}<s_{2}} ds_{1}\, ds_{2}\,  j_{b}^{(0)}(s_{1})\, j_{c}^{(0)}(s_{2})\, . 
\label{Yangianstructure}
\ee
where the level-one generators contain a non-local 
piece related to the non-trivial co-product of \eqn{coproducts}.
The local contributions derive from a conserved current 
$j_{m, a}(s,\tau)$ with 
$\partial^{m}j_{m, a}=0$ ($m=0,1$) in the sense of
$j^{(0)}_{a}(s)=j_{0, a}(s,0)$ and $j^{(1)}_{a}(s)=j_{1, a}(s,0)$. 
Eqs.~\eqn{comrels} are then obeyed at the classical level via Poisson-bracket
relations.

For the case of our interest $Y(\alg{psu}(2,2|4))$ the level-zero generators of the 
superconformal group together with 
one level-one generator, e.g.~the momentum $P^{(1)}_{\mu}$, span
the entire Yangian 
\be
Y(\alg{psu}(2,2|4)) = \text{span} \left ( P^{(0)}_{\mu}, M^{(0)}_{\mu\nu}, K^{(0)}_{\mu}, D^{(0)},
Q_{A}^{\a\, (0)}, \bar Q^{A\,\da\, (0)}\, ; P^{(1)}_{\mu}\, \right ) \, .
\ee

The non-local terms in $J^{(1)}_{a}$ for $Y(\alg{psu}(2,2|4))$ have been 
constructed in a discrete representation
in the context of super-amplitudes
in \cite{Drummond:2009fd}. For this
representation the validity of the super-Serre relations was established  \cite{Dolan:2004ps}.

It is then straightforward to translate the result of \cite{Drummond:2009fd}
 to the continuous case. To begin with
let us focus on the non-local contribution to the
level-one  momentum generator which reads
\begin{align}
P^{(1) \, \mu}_{\text{nl}} = \int \diff s_1  \diff s_2 \: \Bigl\lbrace
\Bigl(\, & d(s_{1})\, \eta^{\mu\nu} - m^{\mu\nu}(s_{1})\, \Bigr)\, p_{\nu}(s_{2})
- \frac{i}{4}  \, {\bar q}^{A\, \da}(s_1)  \, \bar{\sigma}^{\mu}_{\a\da}\,  q_A^{\a}(s_2)
\nn \\ &
 -  
\left( s_1 \leftrightarrow s_2 \right) \Bigr\rbrace \, \theta(s_2-s_1) \, ,
\label{P1full}
\end{align}
where $m^{\mu\nu}(s)$, $p^{\mu}(s)$ and $d(s)$ denote the densities of \eqn{densities1}.
Let us postpone the form of the super-charge densities $q_A^{\a}(s)$ and ${\bar q}^{A\, \da}(s)$ 
for the moment and focus on the bosonic part of the level-one momentum generator
denoted by $P^{(1) \, \mu}_{\text{bos}}$.

Before we embark on the explicit evaluation of $P^{(1) \, \mu}_{\text{bos}}$ acting on 
$\vev{W(C)}_{(1)}$ we need to discuss the regularization of 
\eqn{P1full}. This is necessary as $P^{(1) \, \mu}_{\text{bos}}$ contains two functional
derivatives which may act on the same point along the Maldacena-Wilson loop  giving rise to ill defined terms such as $\delta(0)$. It is natural to introduce
a point-splitting regulator $\epsilon$ by demanding that $s_2 > s_1 + \varepsilon$ holds. 
However, this condition is not reparametrization invariant. For this one rather
performs a point-splitting defined via a cut-off parametrized by $\epsilon$ of the
arc-length via
\begin{align*}
s_{1}<s_{2}-d(s_{2},\epsilon) \quad \text{with} \quad 
\int \limits _{s_2 - d(s_2,\varepsilon)} ^{s_2} \diff s \, \rVert \dot{x}(s) \lVert = \varepsilon \mbox{ .}
\end{align*} 
If one parametrizes the curve by arc-length, i.e.~${\dot x}^{2}=-1$, as we always do in concrete
calculations, this subtlety disappears and one simply has $d(s_{2},\varepsilon) = \varepsilon$.
The regularized level-one momentum generator follows from \eqn{P1full} by replacing
$$
\theta(s_{2}-s_{1}) \to \theta(s_2-s_1-d(s_2,\varepsilon) ) \, .
$$
Of course one may not confine oneself to arc length parametrization  ${\dot x}^{2}=-1$  before 
one has acted with the variational derivatives.
After taking the derivatives one may then set $d(s_2,\varepsilon)=\epsilon$. 
This being understood the bosonic part
of the level-one momentum generator may be written as
\begin{align*}
{P}^{(1) \, \mu} _{\text{bos}, \, \varepsilon} &= \int \diff s_1  \diff s_2 \: \Bigl\lbrace
\Bigl (\,  d(s_{1})\, \eta^{\mu\nu} - m^{\mu\nu}(s_{1})\, \Bigr)\, p_{\nu}(s_{2}) -  
\left( s_1 \leftrightarrow s_2 \right) \Bigr\rbrace \, \theta(s_2-s_1-\varepsilon )  \\
& = \int \diff s_1  \diff s_2 \:
\Bigl (\,  d(s_{1})\, \eta^{\mu\nu} - m^{\mu\nu}(s_{1})\, \Bigr )\, p_{\nu}(s_{2}) \, \Bigl ( \theta(s_2-s_1-\varepsilon) + \theta(s_2-s_1 + \varepsilon)\Bigr) \\
& \quad - \int \diff s_1 \:
(\,  d(s_{1})\, \eta^{\mu\nu} - m^{\mu\nu}(s_{1})\, )\, \int \diff s_2 \:  p_{\nu}(s_{2}) \, ,
\end{align*}
where we have used $\theta(x)=1-\theta(-x)$ in the last step.
The last term in the above factorizes into $(M^{\mu\nu}-D\, \eta^{\mu\nu})\, P_{\nu}$
 and we know already that it annihilates $I_{1 2}$ defined in $(\ref{loop})$. Hence we only need to study the action of the generator
\begin{align}
\tilde{P}^{(1) \, \mu} _{\text{bos}, \, \varepsilon} &=  \int \diff s_1  \diff s_2
\Bigl (\,  d(s_{1})\, \eta^{\mu\nu} - m^{\mu\nu}(s_{1})\, \Bigr )\, p_{\nu}(s_{2}) \, 
\theta(s_2-s_1-
\varepsilon)  \,\, +\, \, ( \varepsilon\to -\varepsilon )
\end{align}
on the vacuum expectation value of the Maldacena-Wilson loop.
Details of this rather tedious calculation may be found in the appendix \ref{app:bosyangian}. 
The final result we obtained reads 
\begin{align}
P^{(1) \, \mu} _{\text{bos}, \, \varepsilon} \vev{W(C)}_{(1)}  & = \frac{\lambda}{16\pi^{2}}\, \bigg \lbrace \frac{1}{6} \int \diff \tau \: \dot{x}^{\mu} (\tau) \left ( \frac{\ddot{x}^{2}}{\dot{x}^{4}} - \frac{(\dot{x}\cdot\ddot{x})^{2}}{\dot{x}^{6}} \, \right )+  \notag  \\
& \quad + 16 \int \diff \tau_1 \diff \tau_2 \: \frac{\dot{x}_1  \dot{x}_2 - |\dot{x}_1| |\dot{x}_2| }{(x_1-x_2)^4}\,(x_1^\mu-x_2^\mu) \, \theta ( \tau_2 - \tau_1 - d(\tau_2,\varepsilon)) \bigg \rbrace \label{pbosW}
\end{align}
suppressing contributions proportional to $\delta(\varepsilon)$ as well as $\cO(\varepsilon)$ terms, see \eqn{res2} in the appendix for the full result.

We hence see that a bi-local expression remains under the action of
$P^{(1) \, \mu} _{\text{bos}, \, \varepsilon}$ which is to be expected as we have not
taken into account the fermionic piece of the level-one momentum generator 
$P^{(1) \, \mu} _{\text{ferm}, \, \varepsilon}$ in \eqn{P1full}.
The natural guess then is that the supersymmetric completion of the bosonic Maldacena-Wilson loop will be invariant under the full Yangian symmetry, as the functional form of the bi-local term
above is that of a fermion-propagator in configuration space.

Hence the fermionic completion of the level-one momentum generator
\begin{align}
P^{(1) \, \mu} _{\text{ferm}, \, \varepsilon} =  - \frac{i}{4}  \int \diff s_1 \diff s_2 \,  \bar{q}^{A \da}(s_{1})\, \bar{\sigma}^{\mu}_{\a\da}\,  q_A^{\a}(s_{2}) \, \Bigl( \theta(s_2 - s_1 -
\varepsilon) - \theta(s_1 - s_2 - \varepsilon)\, \Bigr )  ,
\label{aa1}
\end{align}
acting on the additional fermionic terms in the Maldacena-Wilson loop correlator to be established 
should
cancel the unwanted bi-local term in $(\ref{pbosW})$.

\section{Supersymmetric completion of the Maldacena-Wilson loop}
\label{sec: supsymcomp}

The possibility of a supersymmetric completion of the Maldacena-Wilson loop \eqn{eins} 
was already discussed in the early work on the subject \cite{Drukker:1999zq}. 
Implicitly it was even constructed before that in \cite{Harnad:1985bc} which 
established the super-connection for $\mathcal{N}=1$ super Yang-Mills in 10d. The  
Maldacena-Wilson loop is then a specific light-like path in this higher dimensional
superspace with 10d bosonic base. 
We will need the explicit form of the operator to higher orders in anticommuting 
coordinates. Also we work in a 4d formulation from the outset.

\subsection{Construction of the super Maldacena-Wilson loop}

The construction principle is clear: As we saw the bosonic Maldacena-Wilson loop is invariant
under conformal transformations generated by operators acting as first-order
functional derivatives in the space of bosonic loops $x^{\mu}(s)$. 
In order to supersymmetrize this we need to define paths in superspace.
We choose a full non-chiral superspace parametrized by
\be
x_{\da \a}(s)= \sigma^{\mu}_{\da \a}\, x_{\mu}(s)\, , \quad \theta^{A}_{\a}(s)\, , 
\quad \btheta_{A\, \da}(s)  \, .
\ee
The need for a non-chiral superspace is easy to see: In order to cancel the bi-local contribution
in \eqn{pbosW} by acting with $P^{(1) \, \mu} _{\text{ferm}, \, \varepsilon}$ on a 
fermionic correction to $\vev{W(C)}$ it is clear that this correction
has to be of order $\theta\, \bar\theta$ as the supercharge densities $\bar q$ and $q$
start out as variational derivatives in $\bar\theta$ and $\theta$ respectively. 
Would we consider only a chiral superspace $\{x^\mu(s),\theta^{A}_{\alpha}(s)\}$ then the 
bi-local operator $P^{(1) \, \mu} _{\text{ferm}, \, \varepsilon}$ would only give rise to terms
of order $\theta^{2}$ and the result of the bosonic action in \eqn{pbosW} would not 
receive any purely bosonic corrections.

We then make the ansatz for the super Maldacena-Wilson loop 
\begin{align}
\mathcal{W}(C)&= \frac{1}{N} \mathrm{Tr} \: \mathcal{P} \exp \biggl( i \, I \, [A,\psi,\bar\psi,\phi; 
x,\theta,\btheta ]  \biggr)
\label{ansatz1}
\end{align}
with the exponent $I$ possessing  an expansion in Gra{\ss}mann-odd variables
\begin{align}
I \, [A,\psi,\bar\psi,\phi; 
x,\theta,\btheta]=
\oint_{C} d \tau \, \left( \mathcal{I}_{0} + \mathcal{I}_{1}+ \bar{\mathcal{I}}_{1} + \mathcal{I}_{2m} + \mathcal{I}_{2} + \bar{\mathcal{I}}_{2} +
\cO(\{{\bar\theta}^{i}{\theta}^{3-i}\}) \right) \, .
\label{Iexpand}
\end{align}
Here $\mathcal{I}_{0}$ is the exponent of the usual Maldacena-Wilson loop \eqn{eins}.
Consequently  $\mathcal{I}_{n}$ ($\bar{\mathcal{I}}_{n}$) are extensions of order $n$ in $\theta$ ($\btheta$) and the mixed term $\mathcal{I}_{2m}$ is of order $\theta\btheta$. Terms which contain higher orders of Gra{\ss}mann odd coordinates can be neglected in this context, since corrections to the bosonic result \eqref{pbosW} due to the fermionic part of the level-one momentum generator can only arise from terms of $\left\langle \mathcal{W}(C) \right\rangle_{(1)}$ which are of order $\theta\btheta$ as argued above. To be complete at order two in Gra{\ss}mann odd coordinates we nevertheless will also derive the $\mathcal{I}_{2}$ and $\bar{\mathcal{I}}_{2}$ terms.

To proceed with the construction we note the relevant supersymmetry transformations of the fields generated by $\alg{Q}^{\a}_{A}$ and
$\bar{\alg{Q}}^{A \da}$
\begin{align}
\label{eqn: ftrans1}
&\alg{Q}^{\a}_{A}(A^{\b\db})= 2 \, i \, \eps^{\a\b} \, \bar\psi ^{\db}_{A} \,     &\bar{\alg{Q}}^{A \da} (A^{\b \db}) = - 2 \, i \, \eps^{\da\db} \, \psi ^{ A \b} \\
&\alg{Q}^{\a}_{A}(\bar \phi _{B C}) = \sqrt{2} \, i \, \eps_{A B C D} \, \psi ^{D \a} \,  &\bar{\alg{Q}}^{A \da}(\bar \phi _{B C}) = - \sqrt{2} \, i \, ( \bar \psi ^{\da} _B \, \delta ^A _C - \bar \psi ^{\da} _C \, \delta ^A _B ) \\
&\alg{Q}^{\a}_{A}(\psi ^{B \b}) = \ft{i}{2} \, F^{\a \b}  \, \delta ^B _A + i \, \eps ^{\b \a} \, [\bar \phi _{A C} , \phi ^{B C}] \,   &\bar{\alg{Q}}^{A \da} (\psi ^{B \b}) = - \sqrt{2} \, D^{\b \da} \, \phi ^{A B} \\
&\alg{Q}^{\a}_{A}(\bar \psi ^{\db}_{B}) = - \sqrt{2} \, D^{\db \a} \, \bar \phi _{A B} \,   &\bar{\alg{Q}}^{A \da} (\bar \psi ^{\db}_{B}) = - \ft{i}{2} \, F^{\da \db} \, \delta ^A _B + i \, \eps ^{\da \db} \, [\phi ^{A C} , \bar \phi _{B C}]
\label{eqn: ftrans2}
\end{align}
The construction principle for the supersymmetric Maldacena-Wilson loop is to require that the linearized supersymmetry field-transformations
of the exponent term $I$ may equally well be written as a supersymmetric transformation
of the superpath $\{ x_{\da\a}(s),
\theta^{A}_{\a}(s), \btheta_{A\, \da}(s)\}$. For this we note the
representation of the supersymmetry transformations
$Q^{\a}_{A}$ and $\bar Q^{A\,\da}$ acting in the path superspace 
\begin{align}
Q^{\a}_{A}  &=\int \diff s \, \, q^{\a}_{A}(s)  =\int \diff s \, \Bigl ( - \dfrac{\delta}{\delta \theta^{A}_{\a}(s)}
+ i \, \btheta_{A \da}(s)\, \dfrac{\delta}{\delta x_{\a\da}(s)} \, \Bigr ) \\
{\bar Q}^{A\, \da}  &=\int \diff s \, \, \bar{q}^{A\, \da}(s) =\int \diff s \, \Bigl( \dfrac{\delta}{\delta \btheta_{A \da}(s)}
-i \, \theta^{A}_{\a}(s)\, \dfrac{\delta}{\delta x_{\a\da}(s)} \, \Bigr ) \,.
\end{align}
The exponent $I$ in \eqn{Iexpand} is now constructed in such a fashion to obey the key relations
\begin{align}
\alg{Q}^{\a}_{A} (I) = Q^{\a}_{A} (I)\, \qquad
\bar{\alg{Q}}^{A\, \da} (I) =  {\bar Q}^{A\, \da}(I)\, ,
\label{qqs}
\end{align}
i.e.~the supersymmetry variations of the path equal the supersymmetry variations of the fields.
This guarantees the invariance of the vacuum expectation
value of the Maldacena-Wilson superloop seen
by the following simple argument

\be
0=\vev{ \alg{Q}^{\a}_{A}\, \mathcal{W}(C)}= 
\sfrac{i}{N}\, \vev{\Tr \mathcal{P} \{ e^{i \, I}\, 
\alg{Q}^{\a}_{A}\,I\}} =  \sfrac{i}{N}\, \vev{\Tr \mathcal{P} \{e^{i \, I}\, 
Q^{\a}_{A}\,I\}}
= {Q}^{\a}_{A} \, \vev{ \mathcal{W}(C)} \, ,
\ee
where the zero on the left hand side follows from the invariance of the vacuum state.
 
Up to quadratic order in Gra{\ss}mann variables we find the explicit expressions for the
first few components of the Maldacena-Wilson loop exponent $\mathcal{I}$ in \eqn{Iexpand}
\begin{align}
\mathcal{I}_{0}  & = \ft{1}{2}  \, A^{\b \db} \, \dx_{\b \db} - \ft{1}{2} \, \phi^{CD} \, \bar\eta_{CD} \, |\dx|  \\
\mathcal{I}_{1} & = i \, \theta^{B \b} \, \bpsi^{\db}_{B} \, \dx_{\b \db} +  \sqrt{2} \, i \,
\theta^{C}_{\b} \, \psi^{D \b} \, \bar\eta_{CD} \, |\dx|   \\
\bar{\mathcal{I}}_{1} & =  -  i \, \btheta_{B}^{\db} \, \psi^{B \b} \, \dx_{\b \db}  -  \sqrt{2} \, i \, \btheta_{C \db} \, \bpsi^{\db}_{D} \, \eta^{CD} \, |\dx|    \\
\mathcal{I}_{2} & = - \ft{i}{\sqrt{2}} \, \theta^{C}_{\g} \, \theta^{B \b}  \Bigl(\partial^{\db \g}    \, \bar{\phi}_{CB} \Bigr)  \dx_{\b \db} + \ft{1}{2 \sqrt{2}} \, \theta^{C}_{\b} \, \theta^{D}_{\g} \, F^{\g \b} _{lin} \, \bar\eta_{CD}\, |\dx| +  \sqrt{2} \, i \, \theta^{C}_{\g} \, \dot{\theta}^{B \g} \, \bar{\phi}_{CB}  \\
\bar{\mathcal{I}}_{2} & = -\ft{i}{\sqrt{2}} \, \btheta_{C \dg} \, \btheta_{B}^{\db} \Bigl(\partial^{\b \dg}  \, \phi^{CB} \Bigr) \dx_{\b \db} - \ft{1}{2 \sqrt{2}} \, \btheta_{C \db} \, \btheta_{D \dg}\, F^{\dg \db}_{lin} \, \eta^{CD} \, |\dx| +  \sqrt{2} \, i \, \btheta_{C \dg} \, \dot{\btheta}_{B}^{\dg} \, \phi^{CB}  \\
\mathcal{I}_{2m} & = \ft{1}{4} \, \theta^{B}_{\g} \, \btheta_{B}^{\db} \, F^{\g \b}_{lin} \, \dx_{\b \db} + \ft{1}{4}\, \theta^{B \b} \, \btheta_{B \dg} \, F^{\dg \db}_{lin} \, \dx_{\b\db} + 2 \, i \, \theta^{B}_{\g} \, \btheta_{C \db}  \Bigl( \partial^{\db \g} \, \bar\phi_{B E} \Bigr)  \eta^{C E} \, |\dx| \nn \\ 
&   -\ft{i}{2} \, \theta^{B}_{\g} \, \btheta_{B \dg}  \Bigl(\partial ^{\g \dg} \, \phi^{C D} \Bigr) \bar\eta_{C D}  \, |\dx| + \ft{i}{2} \, \dot{\theta}^{B}_{\b}  \, \btheta_{B \db} \, \phi^{C D} \, \bar\eta_{C D} \, \ft{\dx^{\b \db}}{|\dx|}  - \ft{i}{2} \, \theta^{B}_{\b}  \, \dot{\btheta}_{B \db} \, \phi^{C D} \, \bar\eta_{C D}  \, \ft{\dx^{\b \db}}{|\dx|}  \, .
\end{align}
The details on the computation can be found in appendix \ref{app: Susycompl}. At leading order in the $\theta$ expansion these terms coincide with the results spelled
out in appendix C of \cite{Drukker:1999zq} using a ten dimensional and euclidean notation. 

It is now straightforward to compute the one-loop contribution to the vacuum
expectation value of $\mathcal{W}(C)$.
Using the conventions and propagators in Appendix \ref{app:Propagators}
we find
\begin{align}
\label{susyWilsonOneLoop}
\left\langle \mathcal{W}(C) \right\rangle_{(1)} = -\frac{\lambda}{4 \pi^2}  \int \mathrm{d} \tau_1 \mathrm{d} \tau_2 &\left\{
\left( \frac{1}{4} -  i \Bigl( \btheta_2 \sigma_\mu \theta_1 \Bigr) \frac{x_{12}^\mu}{x_{12}^2} \right) \left( \frac{\dot{x}_1\cdot \dot{x}_2- \left| \dot{x}_1 \right| \left| \dot{x}_2 \right|}{(x_1-x_2)^2} \right) \right.  \nn\\
 & \quad
 + \Bigl( \btheta_2 \sigma_\mu \theta_1 - \btheta_2 \sigma_\mu \theta_2 \Bigr) \frac{\epsilon^{\mu \nu \rho \kappa} \dot{x}_{1\,\nu} \dot{x}_{2\,\rho}x_{12\,\kappa}}{(x_1-x_2)^4} \nn \\
& \quad
+ \frac{i}{2} \left( \btheta_2 \sigma_\mu \dot{\theta}_1 \right) \frac{\dot{x}_2^\mu}{x_{12}^2} - \frac{i}{2} \left( \dot{\btheta}_2 \sigma_\mu \theta_1 \right) \frac{\dot{x}_1^\mu}{x_{12}^2} \nn \\
 & \quad
 \left.-  \frac{i}{2} \left( \btheta_2 \sigma_\mu \dot{\theta}_2 - \dot{\btheta}_2 \sigma_\mu \theta_2 \right) \frac{1}{x_{12}^2}\frac{\left| \dot{x}_1 \right|}{\left| \dot{x}_2 \right|}\dot{x}_2^\mu\right\} \, .
\end{align}
Taking this result we may indeed check the Maldacena-Wilson loop to be supersymmetric 
at one loop
\begin{align}
Q^{\a}_{A} \left\langle \mathcal{W}(C) \right\rangle_{(1)} = 0 && \bar Q^{A\,\da} \left\langle \mathcal{W}(C) \right\rangle_{(1)} = 0 \; .
\end{align}
as is shown in appendix \ref{app:qqbarsym}. 

Full superconformal invariance, especially invariance under $K_{\alpha \dot \alpha}$ at level-zero is, however, not yet expected. 
 The reason for this is that our present choice
of super-space  $\{ x_{\a\da}, \theta^{A}_{\a}, {\bar\theta}_{A\, \da}\}$ lacks the inclusion of bosonic coordinates ${y_A}^B$ for the $R$-symmetry degrees of freedom. As is discussed in appendix \ref{app:non_chiral_psu} the closure of the $\alg{su}(2,2|4)$ algebra requires
this inclusion in the form of derivatives in $y_{A}{}^{B}$ for the representation of generators
$S^{\a}_{A}, {\bar S}^{A\, \da}$ and $K^{\a\da}$. In this respect we have presently constructed 
the exponent $I(x,\theta,\bar\theta, y)$ of the super Maldacena-Wilson loop only for $y_{A}{}^{B}=0$.

\subsection{Yangian invariance of the super Maldacena-Wilson loop at weak coupling} 

Let us now turn to the key question of the potential level-one Yangian invariance of
our result for $\left\langle \mathcal{W}(C) \right\rangle_{(1)}$ in
\eqn{susyWilsonOneLoop}. We recall the result of the action of the level-one momentum
operator in the purely bosonic case in \eqn{pbosW}.
The detailed evaluation of the corrections due to the presence of fermionic terms may be found in appendix \ref{app:Formulas}. The final result we obtained reads
\begin{align}
\left. P^{(1) \, \mu} _{\text{ferm}, \, \varepsilon} \left\langle \mathcal{W}(C) \right\rangle_{(1)} \right|_{\substack{\theta=0 \\ \btheta=0}}  =
- \frac{\lambda}{16\pi^{2}}\, \bigg \lbrace & 16 \int \diff \tau_1 \diff \tau_2 \: \frac{\dot{x}_1 \cdot \dot{x}_2 - |\dot{x}_1| |\dot{x}_2|}{(x_1-x_2)^4}\, (x_1^\mu-x_2^\mu) \, \theta ( \tau_2 - \tau_1 - d(\tau_2,\varepsilon)) \nn\\
& + \frac{4}{3} \int \diff \tau \: \dot{x}^{\mu} (\tau) \left ( \frac{\ddot{x}^{2}}{\dot{x}^{4}} - \frac{(\dot{x}\cdot\ddot{x})^{2}}{\dot{x}^{6}} \, \right ) \bigg \rbrace \, . 
\label{pfermW}
\end{align}
Remarkably, the terms are of the same type as those we encountered in the bosonic result \eqn{pbosW}. In the above we have again dropped terms of order $\delta(\epsilon)$ and
$\cO(\epsilon)$ for the full result see \eqn{here33}\footnote{However, it is worth mentioning that the structure of the neglected $\delta(\varepsilon)$-terms in both parts (i.e. the "bosonic part" and the "fermionic part") of the calculation is the same, but their coefficients do not cancel out.} 
Adding up the two contributions \eqn{pbosW} and \eqn{pfermW} yields the complete result
%
%

%
%
%
\begin{align}
\lim_{\varepsilon \rightarrow 0}\left. P^{(1) \, \mu}_{\text{nl},\, \varepsilon} \left\langle \mathcal{W}(C) \right\rangle_{(1)} \right|_{\substack{\theta=0 \\ \btheta=0}} & = - \frac{7 \, \lambda}{96\pi^{2}}\,\int \diff \tau \:
\left ( \frac{\ddot{x}^{2}}{\dot{x}^{4}} - \frac{(\dot{x}\cdot\ddot{x})^{2}}{\dot{x}^{6}} 
\, \right ) \dot{x}^{\mu} \, ,
\end{align}
where all bi-local terms have canceled out!
What remains after taking the limit $\varepsilon \rightarrow 0$ is a simple  reparametrization invariant curve integral. This in fact defines the \emph{local} contribution to the level-one
momentum generator
\begin{align}
\label{finalwc}
P^{(1) \, \mu}  := P^{(1) \, \mu}_{\text{nl}} +
\frac{7 \, \lambda}{96\pi^{2}}\,  \int \diff \tau \:
\left ( \frac{\ddot{x}^{2}}{\dot{x}^{4}} - \frac{(\dot{x}\cdot\ddot{x})^{2}}{\dot{x}^{6}} 
\, \right ) \dot{x}^{\mu}  \, .
\end{align}
We have thus detected a local contribution to the level-one Yangian generators
at the one-loop order, c.f.~equations \eqn{coproducts} and \eqn{Yangianstructure}.
Then indeed up to this order in perturbation theory and at leading order in the 
$\theta$-expansion we have uncovered a hidden 
symmetry of the appropriately supersymmetrized Maldacena-Wilson loop
\be
P^{(1) \, \mu}\, \vev{\mathcal{W}(C)} = 0 \, .
\ee
We shall now see that this symmetry may be also found at strong coupling.

\section{Strong coupling analysis}
\label{sect:sc}

At strong coupling the expectation value of the Wilson loop is determined by the regularized minimal area in $AdS_5$ \cite{Maldacena:1998im,Rey:1998ik}:
\begin{equation}\label{Loop_at_strong_coupling}
 \mathcal{W}(C)=\,{\rm e}\,^{-\frac{\sqrt{\lambda }A(C)}{2\pi }}.
\end{equation}
The area is computed by minimizing the string action\footnote{We use the standard Poincar\'{e}-patch metric of $AdS_5$ and switch to the Euclidean signature both in target space and on the worldsheet.}:
\begin{equation}\label{A(C)}
 A(C)=\frac{1}{2}\int_{\rm reg}^{}d\tau\, ds \,
 \sqrt{h}h^{ab}\,\frac{1}{Z^2}\left(\partial _aX^\mu \partial _bX_\mu +
 \partial _a Z\partial _bZ\right) - \frac{L(C)}{\epsilon }\,.
\end{equation}
The minimal surface  is
subject to the boundary conditions $Z(0,s )=0$, $X^\mu (0,s)=x^\mu (s )$, where $x^\mu (s)$ parametrizes the contour $C$ on the boundary. In this section we commit ourselves to
the $\dot{x}^2=1$ gauge, and also fix the conformal gauge for the worldsheet metric: $h_{ab}=\delta _{ab}$.

As the area diverges at small $Z$, it has to be regulated by subtracting a boundary counter-term. 
Regularization consists in discarding the slice of the minimal surface with $Z<\epsilon $  for some small $\epsilon $, subtracting a divergent counter-term  proportional to the perimeter   $L(C)$ of the Wilson loop, and then sending $\epsilon $ to zero.

 Integrability of  the string sigma-model in $AdS_5$ guarantees that Wilson loops   satisfy Yangian Ward identities, just because the equations of motion for the minimal surface admit additional conservation laws.
  Following the idea of \cite{Polyakov-unpublished}, we derive
 the Yangian  identities for the minimal area from a higher analogue of the Hamilton-Jacobi equation \footnote{The results in this section were obtained in collaboration with A.~Sever and P.~Vieira.}.

\subsection{Integrability}

\begin{figure}[t]
\begin{center}
\includegraphics[height=6cm]{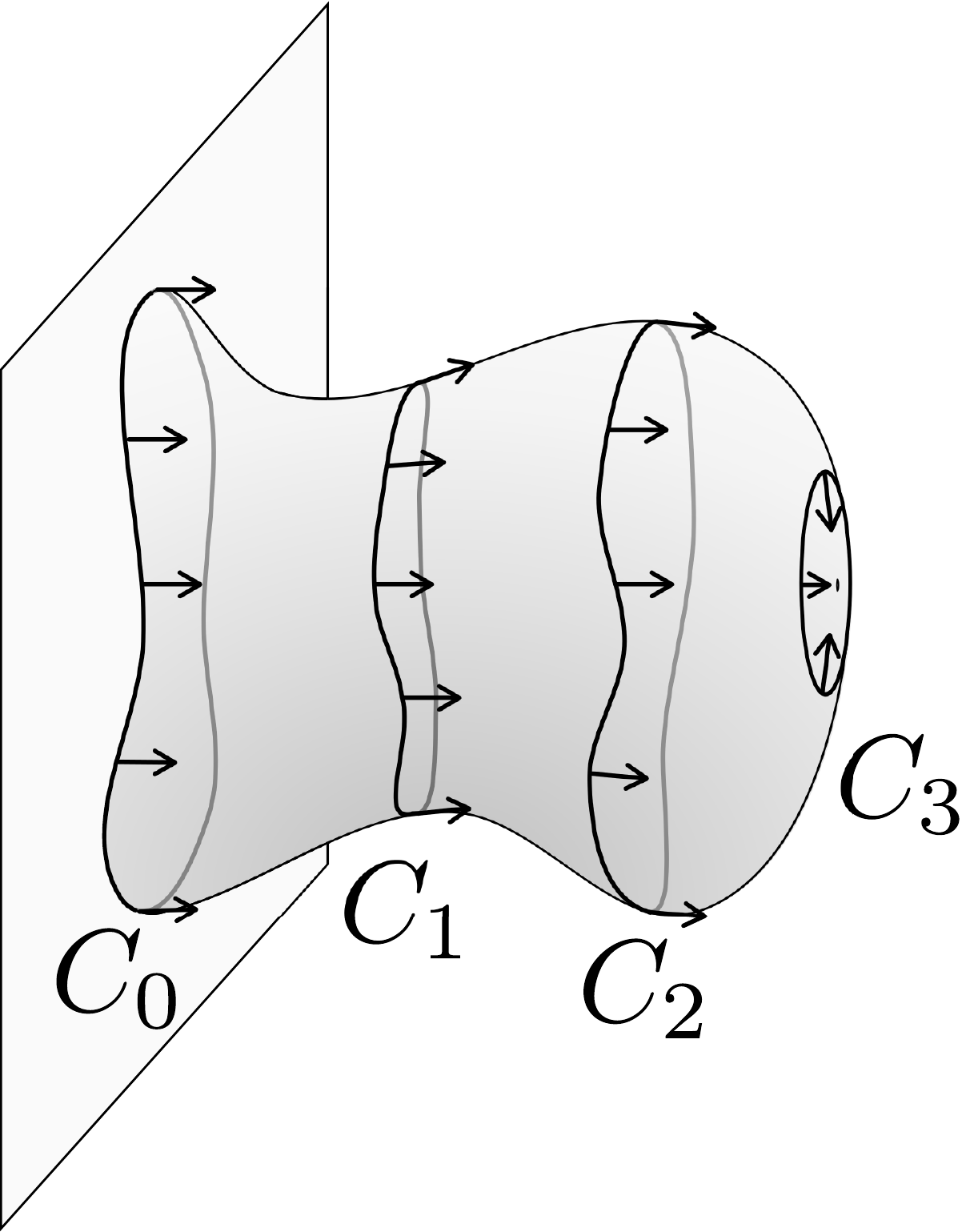}
\caption{As any cycle $C$ on the minimal surface worldsheet is contractible and the charges $Q$
and $Q^{(1)}$ do not change under cycle deformations the initial cycle $C_{0}$  
at the boundary may be shrunk to zero at the tip of the surface. This forces the
charges $Q$ and $Q^{(1)}$ to vanish. }
\label{fig:one}
\end{center}
\end{figure}

 The equations of motion of the sigma-model on $AdS_5$ are equivalent to the conservation law of the isometry current
\begin{equation}\label{current-conservation}
 \partial _iJ^i=0.
\end{equation}
The current takes values in the $\mathfrak{so}(5,1)$ isometry algebra
\begin{equation}\label{concurrents}
 J_i=\partial _i^{\vphantom{i}}X^m\hat{\Xi }_m,\qquad \hat{\Xi }_m=\Xi _{m\,a}\hat{T}^a.
\end{equation}
Here $i=\tau ,s$ are the worldsheet Lorentz indices, $X^m=(X^\mu ,Z)$ are the embedding coordinates of the string, $\Xi _a^m$, $a=1\ldots 15$, are the $AdS_5$ Killing vectors, and $\hat{T}^a$ are the generators of  $\mathfrak{so}(5,1)$ defined in appendix~\ref{conf_algebra:appendix}. Because any cycle on the worldsheet is contractible, the charge associated with the isometry current equals to zero
\begin{equation}\label{charge=0}
 Q\equiv \int_{\tau =\,{\rm const}\,}^{}ds\,J_\tau =0 \, ,
\end{equation}
see figure \ref{fig:one} for a pictorial argument.
 This equation is not an identity and is only valid on-shell, when the embedding coordinates satisfy the equations of motion. We will later derive conformal Ward identities for the minimal area from this equation.

The isometries of $AdS_5$ can be uplifted from the conformal transformations on the boundary. Indeed, if $\xi ^\mu _a$ are the conformal Killing vectors on $\mathbbm{R}^4$ satisfying
\begin{equation}\label{Killing}
 \partial _{(\mu }\xi _{\nu )a}=\frac{1}{2}\,\eta _{\mu \nu }\partial_\lambda \xi^\lambda_a,
\end{equation}
then
\begin{equation}\label{Xi-xi}
 \Xi ^\mu _a =\xi ^\mu_a -\frac{z^2}{8}\,\partial _\mu \partial _\nu \xi ^\nu_a ,\qquad \Xi ^z_a=\frac{z}{4}\,\partial _\mu \xi ^\mu_a. 
\end{equation}
satisfy the Killing equation in the AdS metric
\begin{equation}
 \nabla_{(m}^{\vphantom{i}}\Xi ^i_{n)a}=0.
\end{equation}
Together with the equations of motion for the embedding coordinates, the Killing equation guarantees that the current (\ref{concurrents}) is conserved. 

The explicit form of the conformal Killing vectors can be read off from
\begin{equation}\label{Killing-hat}
 \hat{\xi }^\mu  \equiv \xi ^\mu_a \hat{T}^a
 =\hat{P}^\mu -\hat{M}^{\mu \nu }x_\nu +\hat{D}x^\mu +\hat{K}^\mu x^2-2\hat{K}^\nu x_\nu x^\mu .
\end{equation}
The commutation relations (\ref{[xi,xi]})-(\ref{[dxi,ddxi]}) then imply\footnote{The indices of $\hat{\xi }^\mu $ are raised and lowered with the flat Euclidean metric, while the indices of  $\hat{\Xi }^m$ are transformed with the $AdS_5$ metric.}
\begin{equation}
 \left[\hat{\Xi }_\mu ,\hat{\Xi }_\nu \right]=-\frac{1}{2z^2}\,\partial _{[\mu }\hat{\xi }_{\nu ]},\qquad 
 \left[\hat{\Xi }_\mu ,\hat{\Xi }_z\right]=-\frac{1}{z^3}\,\hat{\xi }_\mu 
 -\frac{1}{8z}\,\partial _\mu \partial _\nu \hat{\xi }^\nu .
\end{equation}
As a consequence of these equations together with (\ref{Xi-xi}), the current (\ref{concurrents}) is not only conserved but is also flat
\begin{equation}\label{flatness}
 \partial _iJ_j-\partial _jJ_i+2\left[J_i,J_j\right]=0.
\end{equation}
The flatness condition is actually an identity, independent of whether the embedding coordinates of the string satisfy the equations of motion or not.

The existence of a flat conserved current is a hallmark of integrability. Such a current implies the existence of an infinite number of conserved charges, local or non-local depending on which basis one chooses. The Yangian symmetry is associated with the non-local charges. The first Yangian charge has the following form
\begin{equation}\label{firstYangian}
 Q^{(1)}=\frac{1}{2}\int_{}^{}ds_1\,ds_2\,\epsilon \left(s_1-s_2\right)
\left[ J_\tau (s_1)J_\tau (s_2)\right]-\int_{}^{}ds\,J_s(s),
\end{equation}
where $\epsilon (s)=\theta (s)-\theta (-s)$ is the anti-symmetric step function, c.f.~our
 discussion in section \ref{sect:Yangiansym}.

Usually, the Yangian charge is conserved only on an infinite line, while on a periodic interval the conservation condition acquires a boundary term. In our  case   the spacial coordinate on the worldsheet is periodic, but it turns out that the boundary term vanishes and the Yangian charge is exactly conserved. This can be understood from the following heuristic argument. A closed Wilson loop can be mapped to an open Wilson line passing through infinity by a global conformal transformation. The spacial coordinate will then have an infinite range and the Yangian charge will be automatically conserved. The conformal transformation that maps a finite loop to an infinite line is actually anomalous \cite{Drukker:2000rr} (see also \cite{Semenoff:2002kk} for the string derivation), but we need not rely on this indirect argument, as the direct computation 
simply shows that the Yangian charge is conserved. Using the current conservation (\ref{current-conservation}), we get
\begin{equation}
 \partial _\tau Q^{(1)}=-\int_{}^{}ds\,\left(\partial _\tau J_s-\partial _sJ_\tau +\left[J_\tau ,J_s\right]\right)-[J_s(L)-J_s(0),Q]=0.
\end{equation}
The bulk term cancels due to the flatness condition (\ref{flatness}), while the boundary contribution vanishes because the isometry charge $Q$ is equal to zero. Therefore the Yangian charge is conserved, and in fact equals to zero
\begin{equation}
 Q^{(1)}=0\, ,
\end{equation}
by the same argument as in eq.~\eqn{charge=0} see again figure \ref{fig:one}.

\subsection{Conformal Ward identities}

As a warm-up exercise we first derive conformal Ward identities for the Wilson loop at strong coupling from conservation of the isometry charge. The derivation relies on the Taylor expansion of the minimal surface near the boundary. Because the AdS metric is singular the first terms in this expansion are completely fixed by the equations of motion, the boundary conditions and the Virasoro constraints \cite{Polyakov:2000ti,Polyakov:2000jg}
\begin{align}\label{XTaylor}
	X^\mu (s,\tau ) &= x^\mu (s)+0\cdot \tau +\frac{1}{2}\,\ddot{x}^\mu (s)\tau ^2-\frac{1}{3}\,p^\mu (s)\tau ^3 \nn \\
 & \quad -\left(\frac{1}{8}\,\ddddot{x}^\mu +\frac{1}{3}\,\ddot{x}^2\ddot{x}^\mu
 +\frac{1}{3}\,\dddot{x}^\nu \ddot{x}_\nu\dot{x}^\mu
 \right)\tau ^4
 +\ldots
 \\
 \label{ZTaylor}
 Z(s,\tau ) &= \tau+0\cdot \tau ^2 -\frac{1}{3}\,\ddot{x}^2(s)\tau ^3+\ldots
\end{align}
The first coefficient that is not fixed by the boundary conditions at $\tau =0$ is $p^\mu (s)$, but it can be related to the variational derivative of the minimal area \cite{Polyakov:2000ti,Polyakov:2000jg}
\begin{equation}
 p^\mu (s)=\frac{\delta A}{\delta x_\mu (s)}\,.
\end{equation}

Plugging the near-boundary expansion of the embedding coordinates into  the isometry current (\ref{concurrents}) we get
\begin{eqnarray}\label{Jt}
 J_\tau &=&\left(\dot{x}^\mu \hat{\xi } _\mu \right)\dot{}\,\,\frac{1}{\tau }-\hat{\xi } ^\mu \,\frac{\delta A}{\delta x^\mu }
 -\left[\frac{1}{24}\,\ddot{x}^2\partial _\mu \hat{\xi } ^\mu 
 +\left(\frac{1}{2}\,\dddot{x}^\mu +\frac{2}{3}\,\ddot{x}^2\dot{x}^\mu \right)
 \!\dot{\vphantom{A}}\,\,\hat{\xi } _\mu 
 \right]\tau 
+O\left(\tau^2 \right)
 \\
 \label{Js}
 J_s&=&\dot{x}^\mu \hat{\xi } _\mu \,\frac{1}{\tau ^2}
 + \left(\dddot{x}^\mu +\frac{2}{3}\,\ddot{x}^2\dot{x}^\mu \right)\hat{\xi }_\mu
 -\left(\frac{1}{2}\,\ddot{x}^\mu \hat{\xi } _\mu  +\frac{1}{8}\,\partial _\mu \hat{\xi } ^\mu \right)\!\dot{\vphantom{A}}+O\left(\tau \right).
\end{eqnarray}
All time-depend terms in $J^i_\tau $ are total derivatives and integrate to zero, as they should. This is just a consequence of charge conservation.  As the total charge vanishes 
the zeroth-order term should also integrate to zero. This gives the constraint
\begin{equation}
 \oint ds\,\xi ^\mu_a \,\frac{\delta A}{\delta x^\mu }=0 \, ,
\end{equation}
being nothing but the conformal Ward identity for the regularized minimal area. We thus formally proved that the minimal area, and with it the Wilson loop 
 at strong coupling are invariant under infinitesimal conformal transformations.
 It is the strong coupling counterpart of our discussion in section \ref{sect:2.1}.

\subsection{Yangian Ward identities}

The Yangian Ward identity is derived in the same way, by expanding the condition  $Q^{(1)}=0$ at small $\tau $. Using (\ref{Jt}), (\ref{Js}) we find at order $O(1/\tau ^2)$ and $O(1/\tau )$
\begin{eqnarray}\label{yang1}
 &&\oint_{}^{}dx^\mu \,\left\{
 \hat{\xi } _\mu +\left[\hat{\xi } _\mu ,\left(\dot{x}^\nu \hat{\xi }_\nu\right)\dot{}\right]
 \right\}=0
 \\
 \label{yang2}
 &&\oint dx^\mu \,
 \left[\hat{\xi }_\mu, \hat{\xi }^\nu\right]\,\frac{\delta A}{\delta x^\nu }=0.
\end{eqnarray}
As expected these equations are  identically satisfied by virtue of eqs.~(\ref{[xi,xi]}) and (\ref{[xi,dxi]}).

A non-trivial equation is obtained at the next order $O(\tau ^0)$ term in $Q^{(1)}$
\begin{equation}\label{Yangian-strong-final}
 \frac{1}{2}\int ds_1\,ds_2\,\epsilon (s_1-s_2)\left[
 \hat{\xi } _{1}^{\mu },\hat{\xi } _{2}^{\nu }\right]\,\frac{\delta A}{\delta x_{1}^{\mu }}\,\,\frac{\delta A}{\delta x_{2}^{\nu }}
- \int_{}^{}ds\,\hat{\xi } ^\mu \left(\ddot{x}^2\dot{x}_\mu +\dddot{x}_\mu \right)=0.
\end{equation}
In the course of the derivation we used the Killing vector identities from the appendix~\ref{conf_algebra:appendix}, which greatly simplify the local term.

Finally, given the minimal area law (\ref{Loop_at_strong_coupling}), we find the that the Wilson loop vacuum expectation value at strong coupling satisfies a second-order variational equation
\begin{equation}
\mathcal{Q}^{(1)}_a \mathcal{W}(C)=0
\end{equation}
with\footnote{An additional second-derivative term that arises upon application of this operator to (\ref{Loop_at_strong_coupling}) has relative order $O(1/\sqrt{\lambda })$ and can be neglected in the $\lambda \rightarrow \infty $ limit.}
\begin{equation}
 \mathcal{Q}^{(1)}_a
 =f^{bc}_a\int ds_1\,ds_2\,\epsilon (s_1-s_2)\xi ^\mu _{1\,b}\xi _{2\,c}^\nu \,\frac{\delta^2}{\delta x_{1}^{\mu }\delta x_{2}^{\nu }}
-\frac{\lambda }{2\pi ^2}\int_{}^{}ds\,\xi_a ^\mu \left(\ddot{x}^2\dot{x}_\mu +\dddot{x}_\mu \right)
\end{equation}
Projected onto the momentum generator, $P^{(1)\, \mu} $, this expression has exactly the same structure as the bosonic part of the Yangian generator at weak coupling \eqn{finalwc}, 
which was written in an ungauged fashion. Taking $\dx^{2}=1$ in \eqn{finalwc}
one recovers the above local term
except for the value of the coefficient of the local term that differs by a factor of $7/24$.

\section{Conclusions and outlook}

In this work we have presented substantial evidence for the existence of a hidden
Yangian symmetry for smooth supersymmetric Maldacena-Wilson loops in $\mathcal{N}=4$ 
SYM theory. For this the level-one generators of the Yangian algebra were shown to 
annihilate the expectation value of the Wilson loop operator at leading order perturbation theory
as well as at leading order in the strong coupling limit upon employing the classical
AdS string description. While the classical AdS string analysis remained purely bosonic,
on the weak coupling gauge theory side it was necessary to consider the supersymmetric 
completion of the original Maldacena-Wilson loop operator of \cite{Maldacena:1998im}. This completion
requires the definition of a loop operator coupling to all the fields of the $\mathcal{N}=4$
multiplet to a path in  an off-shell superspace coordinatized by 
$\{\, x_{\a\da}, \theta^{A}_{\a},{\bar\theta}_{A\, \da}, y_{A}{}^{B}\,\}$. We have 
explicitly constructed this Wilson loop operator
 to quadratic order in anti-commuting coordinates and for $y_{A}{}^{B}=0$.
After computing the one-loop vacuum expectation value of this operator 
the invariance under the action of the Yangian level-one 
momentum generator was established.
Here next to the canonical non-local piece
a local contribution to the Yangian generator appeared. 
Compared to the Yangian generators annihilating super-amplitudes (or light-like 
polygonal super Wilson loops) the emergence of such a local term is novel, although
it does appear in the spectral parameter deformed amplitudes of \cite{Ferro:2012xw,Ferro:2013dga}. 
Consistently the same variational symmetry generators were shown to also annihilate the 
minimal surface at strong coupling. The only difference here is a differing numerical
coefficient in front of the local-piece of the level-one generator. 
It would be
interesting to investigate the light-like limit of our construction and find its
relation to the light-like 
polygonal super Wilson loops \cite{CaronHuot:2010ek,Mason:2010yk,Belitsky:2012nu,Beisert:2012xx}. We note that naively our local term is singular in the light-like
limit. 

A further issue is the above-mentioned coefficient in front of the local term. In general it should
be a function of the coupling constant $\lambda$. Interestingly we find that  in both limits -- at leading
order in weak and strong coupling -- this function
is of order $\lambda$. The two coefficients, however, do not agree. We can offer two 
possible explanations. For one it is natural to expect the existence of 
an interpolating function in front of the local term receiving
corrections to the limits considered.
That function happens to limit to a linear behavior in the weak and strong coupling limits.
On the other hand it is intriguing that opposed to the
weak coupling analysis it was not
necessary at the strong coupling to include the fermionic degrees of freedom.  From the perspective of the IIB superstring in $AdS_{5}\times S^{5}$ a
superpath is actually natural as the superstring ends on the trajectory of a superparticle 
on the boundary. Whether the inclusion of fermionic degrees of freedom on the string side
will affect the purely bosonic local term and its coefficient is unclear to us at this
point. We cannot exclude the possibility that switching on the $y_A^{\hphantom{A}B}$ couplings on the weak-coupling side can also modify the result.
In this context it would be also interesting to explore the consequences of
$\kappa$-symmetry on the boundary.

One may wonder how to interpret a Wilson loop in superspace physically. A good way
to think about it is that the $\theta^{A}_{\a}(s)$ and ${\bar\theta}_{A\, \da}(s)$ parameters
capture the polarizations of a (super)-particle carried along the loop in $\R^{1,3}$.
If one is interested only in the standard Wilson loop one simply projects to the
$\theta^{A}_{\a}(s)={\bar\theta}_{A\, \da}(s)=0$ part. Nevertheless the considered extension may
be useful in establishing new (exact) results on a general $\vev{\mathcal{W}(C)}$.
One could draw a similarity to going super for the on-shell BCFW recursions 
for amplitudes in the theory \cite{ArkaniHamed:2008gz,Brandhuber:2008pf,Elvang:2008na},
which 
led to a complete analytic solution \cite{Drummond:2008cr}
at tree-level.

In any case the hidden Yangian symmetry of Wilson loops should constrain the 
functional form of the vacuum expectation value. It will be important to understand the
structure of invariants and the consequences for possible exact results.  
Finally, the question of how to include a spectral parameter into our considerations
is an obvious one. On the string side equivalence classes of solutions with identical 
regularized areas ending on smooth contours
have been constructed in \cite{Ishizeki:2011bf} parametrized by a spectral parameter. 
For the polygonal light-like situation similar structures were identified in 
\cite{Alday:2010vh}. It is tempting to speculate on a relation to our findings.

\subsection*{Acknowledgments}

We would like to cordially thank N.~Drukker as well as 
A.~Sever and P.~Vieira for independent initial collaborations.
Furthermore we would like to thank S.~Caron-Huot, B.~Eden, S.~Frolov, V.~Kazakov, I.~Kostov, 
and
S.~Vandoren  for important discussions. We are also grateful to 
A.~Sever and P.~Vieira for very useful comments on the draft.
J.~Pl. and K.Z.~thank
the Israel Institute for Advanced Studies in Jerusalem, J.Pl.~and J.Pol.~thank the Kavli IPMU in Tokyo for
hospitality. The work J.Pl. was supported by the Volkswagen-Foundation. The work of K.Z. was supported in part by People Programme (Marie Curie Actions) of the European Union's FP7 Programme under REA Grant Agreement No 317089.

\appendix


\section{Conventions}
\label{app:Conventions}
\subsection*{Minkowski space}
We follow the conventions of \cite{Belitsky:2003sh}. Our signature is $\eta_{\mu\nu}=
\text{diag}(+1,-1,-1,-1)$. Spinor indices are raised and lowered according to 
\begin{align}
\lambda^{\a}&=\epsilon^{\a\b}\, \la_{\b}\, , \qquad \la_\a = \la^{\b} \eps_{\b\a}\, , \qquad
\lambda_{\da}=\epsilon_{\da\db}\, \la^{\db}\, , \qquad \la^\da = \la_{\db} \eps^{\db\da}\, \nn \\
\eps^{12}&=\eps_{12}=1 \, , \quad \eps^{\dot{1}\dot{2}}=\eps_{\dot{1}\dot{2}}=-1 \quad \Rightarrow \eps^{\a\b}\, \eps_{\g\b}= \delta^{\a}_{\g} \, , \quad \eps^{\da\db}\, \eps_{\dg\db}= \delta^{\da}_{\dg} 
\end{align}
We note 
\begin{align}
\sigma^{\mu\, \da\beta}=(\mathbf{1},\mathbf{\sigma}) \, , \qquad {\bar\sigma}^{\mu}_{\a\db}=(\mathbf{1},-\mathbf{\sigma}) \label{sigma}
\end{align} 
with the vector $\sigma$ of Pauli matrices
\begin{align*}
\sigma^0 = \begin{pmatrix} 1 & 0 \\ 0 & 1\end{pmatrix} \quad  \sigma^1 = \begin{pmatrix} 0 & 1 \\ 1 & 0 \end{pmatrix} \quad \sigma^2 = \begin{pmatrix} 0 & -i \\ i & 0 \end{pmatrix} \quad \sigma^3 = \begin{pmatrix} 1 & 0 \\ 0 & -1  \end{pmatrix} \, .
\end{align*} 
If not stated otherwise the index position of the matrices $\sigma$ and $\bar{\sigma}$ is given by $(\mathrm{\ref{sigma}})$. They can be identified as follows: 
\begin{align*}
\sigma^{\mu\, \da\b} = \epsilon^{\b\gamma}\, {\bar \sigma}^{\mu}_{\gamma\dot\delta}\,
\epsilon^{\dot\delta\da} = {\bar\sigma}^{\mu \, \b\da} \qquad
{\bar\sigma}^{\mu}_{\a\db} = \epsilon_{\db\dot\gamma}\, { \sigma}^{\mu\, \dot\gamma\delta}\, \epsilon_{\delta\a}\, = \sigma^\mu_{ \db \a} \, .
\end{align*}
Contractions of space-time or spinor indices give the following results:
\begin{align}
\bar{\sigma}^{\mu}_{\a \db}  \bar{\sigma}_{\mu \, \g \dot\delta} = - 2 \eps_{\a \g} \eps_{\db \dot\delta} \, , \quad \sigma^{\mu}_{\da \b}  \sigma_{\mu \dg \delta} = - 2  \eps_{\b \delta} \eps_{\da \dg} \, , \quad  \bar{\sigma}^{\mu}_{\a \da}\bar{\sigma}^{\nu \, \a \da} = 2 \eta^{\mu \nu} = \sigma^{\mu}_{\da \a} \sigma^{\nu \, \da \a} \label{sigmaid}
\end{align} 
To a space-time vector $p^{\mu}$ we assign a bi-spinor as follows:
\begin{align*}
p^{\a \da} := \bar{\sigma}^{\mu \, \a \da} p_{\mu} =  \sigma^{\mu \, \da \a} p_{\mu} =:   p^{\da \a}
\end{align*} 
The identities $(\mathrm{\ref{sigmaid}})$ imply that
\begin{align*}
p^{\mu} = \half \, \bar{\sigma}^{\mu \, \a \da} p_{\a \da} = \half \, \sigma^{\mu \, \da \a} p_{\da \a} \, \quad \mbox{and} \quad p_{\a \da} k^{\a \da} = 2 \, p_{\mu} k^{\mu} \, .
\end{align*}
Defining
\begin{align*}
\sigma^{\mu \nu \, \a \b} := \ft{i}{2} \left(\bar{\sigma}^{\mu \, \a \dot\delta} \sigma^{ \nu \, \dg \b} - \bar{\sigma}^{\nu \, \a \dot\delta} \sigma^{ \mu \, \dg \b}  \right) \eps_{\dg \dot\delta} \, , \qquad \bar{\sigma}^{\mu \nu \, \da \db} := \ft{i}{2} \left(\sigma^{ \mu \, \da \g} \bar{\sigma}^{\nu \, \delta \db}  - \sigma^{ \nu \, \da \g} \bar{\sigma}^{\mu \, \delta \db}  \right) \eps_{\delta \g} \, ,
\end{align*} 
we also assign bi-spinors to an antisymmetric 2-tensor $F_{\mu \nu}$:
\begin{align}
F^{\a \b} := F_{\mu \nu} \sigma^{\mu \nu \, \a \b} \, , \qquad F^{\da \db} := F_{\mu \nu} \sigma^{\mu \nu \, \da \db} \, .
\label{eqn: Fmunubispin}
\end{align}
These two bi-spinors associated to $F_{\mu \nu}$ can be related to $F^{\a \da \b \db} : = F_{\mu \nu} \bar{\sigma}^{\mu \, \a \da} \bar{\sigma}^{\nu \, \b \db}$ by the following identity:
\begin{align}
F^{\a \da \b \db} = \ft{i}{2} \eps^{\da \db} F^{\a \b} + \ft{i}{2} \eps^{\a \b} F^{\da \db} 
\label{eqn: F4decomp2}
\end{align}
The bi-spinors associated to $F_{\mu \nu}$ are symmetric, $ F^{\a \b} = F^{\b \a} \, , F^{\da \db} = F^{\db \da}$. For other bi-spinors we have the general symmetry property:
\begin{align}
\label{eqn: weylindpartition}
\Lambda_{(\a \b)} &= \Lambda_{\a \b} + \half \eps_{\a \b} \, \Lambda \indices{^{\g}_{\g}} \\
\Lambda_{(\da \db)} &= \Lambda_{\da \db} + \half \eps_{\da \db} \, \Lambda \indices{_{\dg}^{\dg}} 
\end{align} 
We note the Fierz identity
\begin{align}
\bar\xi^{\da}\, \xi^{\b} &= \half \, \sigma^{\mu\, \da\b} \, ( \bar\xi ^{\dg}  \sigma_{\mu \, \dg \delta}\, \xi^{\delta} )\, ,
\end{align}
and some trace identities for the sigma matrices:
\begin{align}
\half \Tr (\bar{\sigma}^{\mu} \sigma^{\nu}) &= \eta^{\mu \nu} \, , \\
\half \, \Tr(\bar\sigma^{\mu}\, \sigma^{\nu}\, \bar\sigma^{\rho}\,\sigma^{\kappa}) &=
\eta^{\mu\nu}\, \eta^{\rho\kappa} + \eta^{\nu\rho}\, \eta^{\mu\kappa} -
\eta^{\mu\rho}\, \eta^{\nu\kappa} -i\,\epsilon^{\mu\nu\rho\kappa} \, , \\
\half \, \Tr(\sigma^{\mu}\, \bar\sigma^{\nu}\, \sigma^{\rho}\,\bar\sigma^{\kappa}) &=
\eta^{\mu\nu}\, \eta^{\rho\kappa} + \eta^{\nu\rho}\, \eta^{\mu\kappa} -
\eta^{\mu\rho}\, \eta^{\nu\kappa} +i\,\epsilon^{\mu\nu\rho\kappa} \, ,
\end{align}
Gra{\ss}mann functional derivatives are defined by:
\begin{align}
 \dfrac{\delta}{\delta \theta^{A}_{\a}(s)} \, \theta^{B}_{\b}(\tau) = \delta ^B_A \, \delta ^{\a}_{\b} \, \delta(\tau-s) \,,  \quad \dfrac{\delta}{\delta \bar{\theta}_{A\, \da}(s)} \, \bar{\theta}_{B \, \db}(\tau) = \delta ^A_B \, \delta ^{\da}_{\db} \, \delta(\tau-s) \, .
\end{align}
\subsection*{Six-dimensional space}
Consider the vector space $\mathbb{R}^6$ with the metric $\eta_{I J} = \diag(-1,\ldots, -1)$. To a vector $\phi_I \in \mathbb{R}^6$ we assign $(4 \times 4)$-matrices by the prescription
\begin{align}
\phi^{A B} := \ft{1}{\sqrt{2}} \Sigma ^{I \, A B} \phi_I \, , \qquad \bar{\phi}_{A B} := \ft{1}{\sqrt{2}} \bar{\Sigma} ^I _{A B} \phi_I \, .
\end{align}
The sigma matrices are given by
\begin{align}
(\Sigma^{1 \, A B}, \ldots , \Sigma^{6 \, A B} ) &= (\eta_{1 \, A B}, \eta_{2 \, A B}, \eta_{3 \, A B}, i \bar{\eta}_{1 \, A B}, i \bar{\eta}_{2 \, A B}, i \bar{\eta}_{3 \, A B}) \, ,  \\
(\bar{\Sigma}^1 _{ A B}, \ldots , \bar{\Sigma}^6 _{ A B} ) &= (\eta_{1 \, A B}, \eta_{2 \, A B}, \eta_{3 \, A B}, - i \bar{\eta}_{1 \, A B}, - i \bar{\eta}_{2 \, A B},- i \bar{\eta}_{3 \, A B}) \, , \\
\eta_{i \, A B} :&= \epsilon_{i A B 4} + \delta_{i A} \delta_{4 B} - \delta_{i B} \delta_{4 A} \, \quad \bar{\eta}_{i \, A B} := \epsilon_{i A B 4} - \delta_{i A} \delta_{4 B} + \delta_{i B} \delta_{4 A} \, .
\end{align}
The epsilon-tensor has the following contraction:
\begin{align}
\eps_{D A B C} \, \eps^{D K L M} = \delta_{A B C} ^{K L M} + \delta_{A B C} ^{M K L} + \delta_{A B C} ^{L M K} - \delta_{A B C} ^{L K M} - \delta_{A B C} ^{M L K} - \delta_{A B C} ^{K M L} \, .
\label{eqn: epseps6delta}
\end{align}
We note the following identities:
\begin{align}
\bar{\phi}_{A B} = \half \eps_{A B C D} \phi^{C D} \, \qquad
\phi^{A B} = \half \eps^{A B C D} \bar{\phi}_{C D} \, \qquad X^{A B} \bar{Y}_{A B} = - 2 X^I Y_I \, .
\end{align} 
For a unit vector $n \in \mathbb{R}^6 \, , n^I n_I = -1 $, we have:
\begin{align}
\bar{n}_{AB} n^{CB} = \half  \, \delta^{C}_{A} \, , \quad \bar{n}_{AB} n^{AB} =2\, .
\end{align}
Note also that our conventions imply that
\begin{align}
\dfrac{\partial}{\partial x _{\a \da}} \, x_{\b \db}  = 2 \, \delta ^\a
_\b \, \delta ^\da _\db \, .\label{spinorderivative}
\end{align}

\section{Propagators}
\label{app:Propagators}

We have with
\begin{align*}
A_{\mu}(x) =  A_{\mu}^a(x) T^a \quad \quad \phi^{a\, AB} (x) = \phi^{a\, AB} (x) T^a \qquad
\psi^{a\, A}_{\alpha} (x) = \psi^{A}_{\alpha} (x) T^a 
\end{align*} 
and $\mathrm{Tr}(T^a T^a) = \half \delta^{a b}$ the propagators 
\begin{align}
\left\langle \phi^{a\,AB}(x_1) \phi^{b\,CD}(x_2) \right\rangle &= - \frac{g^2}{4 \pi^2} \frac{\epsilon^{ABCD} \delta^{ab}}{(x_1-x_2)^2} \\
\left\langle \psi^{a\,A}_\a(x_1) \bar{\psi}^{b}_{\da \, B}(x_2) \right\rangle &= \frac{i g^2}{2 \pi^2} \delta^{ab} \delta^A_B \frac{(x_1-x_2)_{\a \da}}{(x_1-x_2)^4} \\
\left\langle A^{a}_\mu(x_1) A^{b}_\nu(x_2) \right\rangle &= \frac{g^2}{4 \pi^2} \frac{\eta_{\mu \nu} \delta^{ab}}{(x_1-x_2)^2}
\end{align}

\section{Conformal algebra and Killing vectors}\label{conf_algebra:appendix}

We use the conventions of \cite{Drummond:2009fd} for the conformal algebra.
The generators 
$ \left\{P_\mu ,M_{\mu \nu },D,K_\mu \right\}$ are collectively denoted by $T_a$, and satisfy the commutation relations:
\begin{eqnarray} 
 \begin{aligned} 
  \left[ M_{\mu \nu}, M_{\rho \sigma} \right] &=   \eta_{\mu \sigma} M_{\nu \rho} + \eta_{\nu \rho} M_{\mu \sigma} - \eta_{\mu \rho} M_{\nu \sigma} - \eta_{\nu \sigma} M_{\mu \rho}   \\ 
\left[ P_{\mu}, P_{\nu} \right] &= 0 \hspace{40.05mm} \left[ M_{\mu \nu}, P_{\lambda} \right] = \eta_{\nu \lambda} P_{\mu}    -  \eta_{\mu \lambda} P_{\nu} \\   
 \left[ D, P_{\mu} \right] &=  - P_{\mu} \hspace{37.55mm}  \left[ D, K_{\mu} \right] =  K_{\mu} \\ 
\left[ D, M_{\mu \nu} \right] &= 0 \hspace{39.45mm}  \left[ M_{\mu \nu}, K_{\rho} \right] = \eta_{\nu \rho} K_{\mu}   -  \eta_{\mu \rho} K_{\nu} \\  
\left[ P_{\mu}, K_{\nu} \right] &=  2 M_{\mu \nu} -  2 \eta_{\mu \nu} \,  D  \hspace{17.7mm} 
[K_{\mu}, K_{\nu}] = 0 . \end{aligned}   \label{conf_algebra}
\end{eqnarray}
In  the standard $M_{MN}$ basis of $\mathfrak{so}(4,2)$, where $M,N=\mu ,4,4'$ with $\eta _{44}=1=-\eta _{4'4'}$,
\begin{equation}
 D=M_{44'}\qquad P_\mu =M_{\mu 4'}+M_{\mu 4}
 \qquad K_\mu =M_{\mu 4'}-M_{\mu 4}.
\end{equation}
The Killing metric on the $\mathfrak{so}(4,2)$ algebra is
\begin{equation}
\left\langle M_{MN},M_{KL}\right\rangle=\eta_{LM} \eta_{NK} - \eta_{LN} \eta_{MK}.
\end{equation}

The dual basis of generators is defined by raising the indices with the inverse of the Killing metric
\begin{equation}
 \hat{T}^a=G^{ab}T_a,
\end{equation}
such that
\begin{equation}
 \left\langle \hat{T}^a,T_b\right\rangle=\delta ^a_b.
\end{equation}
The structure constants that appear in the Yangian generators are those that arise in the commutation relations in the dual basis:
\begin{equation}
 \left[\hat{T}^a,\hat{T}^b\right]=f^{ab}_c\hat{T}^c.
\end{equation}
Explicitly,
\begin{equation}
 \hat{M}^{\mu \nu }=-\eta ^{\mu \lambda }\eta ^{\nu \rho }M_{\lambda \rho }\qquad 
 \hat{P}^\mu = - \frac{1}{2}\,\eta ^{\mu \nu }K_\nu
 \qquad 
 \hat{K}^\mu = - \frac{1}{2}\,\eta ^{\mu \nu }P_\nu
 \qquad 
 \hat{D}=D,
\end{equation}
and
\begin{eqnarray} 
 \begin{aligned} 
  \left[ \hat{M}^{\mu \nu}, \hat{M}^{\rho \sigma} \right] &=  
  \eta^{\mu \rho } \hat{M}^{\nu \sigma } + \eta^{\nu \sigma } \hat{M}^{\mu \rho } 
   -\eta^{\mu \sigma } \hat{M}^{\nu \rho } - \eta^{\nu \rho } \hat{M}^{\mu \sigma }  
   \\ 
\left[ \hat{P}^{\mu}, \hat{P}^{\nu} \right] &= 0 \hspace{40.05mm} \left[ \hat{M}^{\mu \nu}, \hat{P}^{\lambda} \right] =    \eta^{\mu \lambda} \hat{P}^{\nu} -\eta^{\nu \lambda} \hat{P}^{\mu}\\   
 \left[ \hat{D}, \hat{P}^{\mu} \right] &= \hat{P}^{\mu} \hspace{40.55mm}  \left[ \hat{D}, \hat{K}^{\mu} \right] = -\hat{K}^{\mu} \\ 
\left[ \hat{D}, \hat{M}^{\mu \nu} \right] &= 0 \hspace{39.45mm}  \left[\hat{M}^{\mu \nu}, \hat{K}^{\rho} \right] =  \eta^{\mu \rho} \hat{K}^{\nu} 
-\eta^{\nu \rho} \hat{K}^{\mu}  \\  
\left[ \hat{P}^{\mu}, \hat{K}^{\nu} \right] &= \frac{1}{2}\, \eta^{\mu \nu}  \hat{D}  -\frac{1}{2}\ \hat{M}^{\mu \nu}\hspace{13.3mm} 
\left [\hat{K}^{\mu}, \hat{K}^{\nu} \right ] = 0 . \end{aligned}   \label{dual_conf_algebra}
\end{eqnarray}

The conformal Killing vectors $\xi ^\lambda  _a=\left\{
\delta ^\lambda  _\mu,x_{\mu }\delta ^\lambda _{\nu }-x_{\nu }\delta ^\lambda _{\mu },x^\lambda ,
x^2\delta ^\lambda _{\mu }-2x^\lambda x_\mu  
\right\}$ satisfy the commutation relations of the conformal algebra,
\begin{equation}
 \left\{\xi _a,\xi _b\right\}=f_{ab}^c\xi _c,
\end{equation}
with respect to the Lie bracket
\begin{equation}
 \left\{u,v\right\}^\mu =u^\nu \partial _\nu v^\mu -v^\nu \partial _\nu u^\mu .
\end{equation}
The following identities for the Lie-algebra-valued Killing vector (\ref{Killing-hat}): 
\begin{eqnarray}\label{[xi,xi]}
 &&\left[\hat{\xi }^\mu ,\hat{\xi }^\nu \right]=0
 \\
 \label{[xi,dxi]}
 &&\left[\hat{\xi }^\mu ,\partial _\nu \hat{\xi }^\lambda \right]=
 \eta ^{\mu \lambda }\hat{\xi }_\nu -\delta ^\mu _\nu \hat{\xi }^\lambda 
 -\delta ^\lambda _\nu \hat{\xi }^\mu 
 \\
 &&\left[\hat{\xi }^\mu ,\partial _\nu \partial _\lambda \hat{\xi }^\rho \right]=\eta _{\nu \lambda }\partial ^\rho \hat{\xi }^\mu 
 -\delta ^\rho _\nu \partial _\lambda \hat{\xi }^\mu -
 \delta ^\rho _\lambda \partial _\nu \hat{\xi }^\mu 
 \\
 \label{[dxi,ddxi]}
 &&\left[\partial _\nu \hat{\xi }^\nu ,\partial _\mu \partial _\lambda \hat{\xi }^\lambda \right]=-4\partial _\mu \partial _\nu \hat{\xi }^\nu .
\end{eqnarray} 
are used in checking the flatness condition of the worldsheet isometry current.

\section{Superconformal algebra in non-chiral superspace}
\label{app:non_chiral_psu}
In the study of superamplitudes in $\mathcal{N} = 4$  SYM one encounters a representation of 
the $\mathfrak{su}(2,2|4)$ superalgebra in the form of differential operators  on a
chiral-superspace  $\{x_{\alpha \dot\alpha}, {\theta_\alpha}^A\}$ \cite{Drummond:2008vq}. Extending this algebra naively into a non-chiral superspace $\{x_{\alpha \dot\alpha}, {\theta_\alpha}^A, \bar\theta_{A \dot\alpha}\}$, where one simply conjugates the $\theta$ expressions to $\bar\theta$'s 
does not work since the resulting algebra will not close. From a coset construction \cite{Howe:1996rk} of the algebra it is clear that we need to introduce coordinates carrying R-symmetry indices ${y_A}^B$. Defining a set of coordinates and their variation under the algebra
\begin{align}
X = \{x_{\alpha \dot\alpha}, {\theta_\alpha}^A, \bar\theta_{A \dot\alpha}, {y_A}^B\}, ~~~~~~~~~~~~~
\label{eqn:su_trafo}
\delta X = \{ \delta x_{\alpha \dot\alpha}, \delta {\theta_\alpha}^A, \delta \bar\theta_{A \dot\alpha}, \delta {y_A}^B\}\; ,
\end{align}
we can understand the superspace in which our Wilson loop is constructed as a subsurface given by putting ${y_A}^B = 0$. But starting at a specific point on the constraint surface
\begin{align}
X_0 = \{(x_0)_{\alpha \dot\alpha}, {(\theta_0)_\alpha}^A, (\bar\theta_0)_{A \dot\alpha}, 0\}
\end{align}
general superconformal variations $\delta X$  will yield $ y_{A}{}^{B} \ne 0$. Therefore not all transformations are expected to be symmetries of our super Wilson loop.

The superconformal transformations (\ref{eqn:su_trafo}) are generated by the following operators:
\begin{align}
M_{\a \b} &= 2 i \, x_{\dg (\a} \partial \indices{_{\b)} ^\dg} + 4 i \, \theta^A _{(\a} \partial_{\b) \, A}  \hspace{20.5mm} \overline{M}_{\da \db} = 2 i \, x \indices{^\g _{(\da}} \partial _{\db) \g} - 4 i \, \btheta _{A\, (\da} \partial_{\db)} ^A \\
D &= \half x_{\a \da} \partial^{\a \da} + \half \theta ^B _\b \partial ^\b _B + \half \btheta_{B \, \db} \partial ^{B \, \db} \hspace{10mm} P^{\a \da} = \partial^{\a \da} \\
K_{\a \da} &= - x_{\a \dg} x_{\da \g} \partial^{\g \dg} - 2 x_{\da \g} \theta^C _\a \partial ^\g _C - 2 x_{\a \dg} \btheta_{C \, \da} \partial ^{C \, \dg} + 4 i \theta ^A _\a \btheta _{B \, \da} \partial \indices{_A ^B} \\
Q_A ^\a &= - \partial ^\a _A + y \indices{_A ^B} \partial _B ^\a + i \btheta_{A \, \da} \partial ^{\a \da} \hspace{18mm}  \overline{Q}^{A \, \da} = \partial ^{A \, \da} + y \indices{_B ^A} \partial ^{B \, \da} - i \theta ^A _\a \partial ^{\a \da} \\
S^A _\a &= \left( \delta ^A _B + y \indices{_B ^A} \right) \left( x_{\a \dg}  \partial ^{B \, \dg} + 2i \theta ^C _\a \partial \indices{_C ^B} \right) - i x_{\a \dg} \theta ^A _\b \partial^{\b \dg} - 2i \theta ^A _\b \theta ^C _\a \partial ^\b _C \\
\overline{S}_{A \, \da} &=\left( - \delta ^B _A + y \indices{_A ^B} \right)\left( x_{\da \g}  \partial ^\g _B - 2i \, \btheta _{D \, \da} \partial \indices{_B ^D} \right) + i x_{\da \g} \btheta_{A \, \db} \partial ^{\g \db} + 2i \btheta_{A \, \db} \btheta_{C \, \da} \partial ^{C \, \db} \\
R \indices {^{\prime \, A} _B} &= 2i \left( - \delta ^D _B + y \indices{_B ^D} \right) \left( \delta ^A _C + y \indices{_C ^A} \right) \partial \indices{_D ^C} + 2i \left(-\delta ^C _B + y \indices{_B ^C} \right) \theta ^A _\g \delta ^\g _C  \nn \\
& \quad  + 2i \left( \delta ^A _C + y \indices{_C ^A} \right) \btheta _{B \, \da} \partial ^{C \, \da} + 2 \btheta_{B \, \da} \theta ^A _\a  \partial ^{\a \da} \nn \\
R \indices {^{A} _B} &= R \indices {^{\prime \, A} _B} - \ft{1}{4} \, \delta^A _B \, R \indices {^{\prime \, C} _C} \\
C&= \ft{1}{4} \big( \theta ^D _\a \partial ^\a _D - \btheta _{C \, \da} \partial ^{C \, \da} + i \theta^A _\a \btheta_{A \, \da} \partial^{\a \da} - \partial \indices{_A ^A} \nn \\
& \quad + y \indices{_A ^B} \theta ^A _\a \partial ^\a _B + y \indices{_A ^B} \btheta_{B \, \da} \partial^{A \, \da} + y\indices{_A ^C} y \indices{_C ^B} \partial \indices{_B ^A}  \big)
\end{align}
where we used the shorthand notation
\begin{align}
\partial^{\dot \alpha \alpha} = \frac{\partial}{\partial x_{\alpha \dot \alpha}} \, ,~~~~~~ & ~~~~~~ \partial^\alpha_A = \frac{\partial}{\partial \theta^A_\alpha} \, ,&
\partial^{\dot \alpha A} = \frac{\partial}{\partial \btheta_{A \dot \alpha}} \, , ~~~~~~ & ~~~~~~  {\partial_A}^B= \frac{\partial}{\partial {y_B}^A}  \; .
\end{align}
Note also that our conventions imply \eqn{spinorderivative}. The commutation relations of the above generators agree with \eqn{conf_algebra} if the generators of the conformal algebra are translated as defined in appendix \ref{app:Conventions}.

The nontrivial part of the algebra, which is realized by these generators is given by
\begin{align}
\Big \{ Q_A ^\a, {\bar Q}^{B \dot \alpha} \Big \} & = 2 i \delta_A^B P^{\dot \alpha \alpha} \nn ~ & ~
\Big \{ S^A_\alpha , {\bar S}_{B \dot \alpha}  \Big \} &= -2 i \delta_B^A K_{\alpha \dot \alpha} \nn \\
\Big[ K_{\gamma \dot  \gamma}, Q_A ^\a \Big] &= 2 \delta_\gamma^\alpha {\bar S}_{A \dot \gamma} \nn  ~ & ~
\Big[K_{\gamma \dot  \gamma}, {\bar Q}^{B \dot \alpha} \Big] &= 2 \delta_{\dot \gamma}^{\dot \alpha} S^B_{\gamma} \nn \\
\Big[ P_{\alpha \dot \alpha} , K_{\beta \dot \beta} \Big] &= i \eps_{\a \b}\, \overline{M}_{\da \db} + i \eps_{\da \db}\,  M_{\a \b} + 4 \, \eps_{\a \b} \, \eps_{\da \db} \, D \\
\Big \lbrace Q^\a _A \, , \,  S^B _\b  \Big \rbrace &= \delta ^B _A \, M \indices{^\a _\b} + \delta ^\a _\b \, R \indices{^B _A} + 2i \, \delta ^B _A \, \delta ^\a _\b \left(D + C \right)  ~ & ~ \Big[ P^{\dot \alpha \alpha}  , S^A_\beta \Big] &= 2 \delta_\beta^\alpha {\bar Q}^{A \dot \alpha} \nn \\
\Big\{ {\bar Q}^{A \dot \alpha}, {\bar S}_{B \dot \beta} \Big\} &= - \delta ^A _B \, \overline{M} \indices{^\da _\db} - \delta ^\da _\db \, R \indices{ ^A _B} + 2i \, \delta ^A _B \, \delta ^\da _\db \left(D - C \right) ~ & ~ \Big[ P^{\dot \alpha \alpha}  , {\bar S}_{A \dot \alpha} \Big] &= 2 \delta_{\dot\beta}^{\dot\alpha} Q_A ^\a  \nn \; .
\end{align}

Making contact with our smaller superspace where $y_{A}{}^{B} = 0$, we see that $P$, $M$, $\overline M$, $Q$ and $\bar Q$ are well defined generators in the sense that they preserve
the constraint surface $y_{A}{}^{B} = 0$.
The remaining generators $K, S, \bar S, R$ contain derivatives w.r.t.~$y_{A}{}^{B}$ and 
are therefore not expected to be symmetries of a constrained $y_{A}{}^{B} = 0$ Wilson loop.
Note that the level-one momentum generator (\ref{P1full}) is also well defined since it only depends on combinations of $P$, $M$, $\overline M$, $Q$ and $\bar Q$.

\section{Details of the weak coupling computation}
\label{app:bosyangian}
\subsection{Non-local variation of \texorpdfstring{$\vev{W(C)}$}{<W(C)>}}
In this appendix we calculate
\begin{align*}
P^{(1) \, \mu} _{\text{bos}, \, \varepsilon} \vev{W(C)}_{(1)}  = - \frac{\lambda}{16 \pi^2} P^{(1) \, \mu} _{\text{bos}, \, \varepsilon} \int \diff \tau_1 \diff \tau_2 \, I_{12} \, .
\end{align*}
For computational purposes it is helpful to rewrite the generator in the following form:
\begin{align*}
P^{(1) \, \mu} _{\text{bos}, \, \varepsilon}  &= \int \diff s_1  \diff s_2
(\,  d(s_{1})\, \eta^{\mu\nu} - m^{\mu\nu}(s_{1})\, )\, p_{\nu}(s_{2}) \left( \theta(s_2-s_1-d(s_2,\varepsilon) + \theta(s_2-s_1 + d(s_1,\varepsilon)\right) \\ 
& =  \int \diff s_1 \diff s_2  N^{\mu \nu \rho \sigma} x_{\nu}(s_1) \dfrac{\delta}{\delta x^{\rho}(s_1)} \dfrac{\delta}{\delta x^{\sigma}(s_2)}  \left( \theta(s_2-s_1-d(s_2,\varepsilon) + \theta(s_2-s_1 + d(s_1,\varepsilon)\right)  ,\\
 N^{\mu \nu \rho \sigma} :&= \eta ^{\mu \sigma} \eta ^{\nu \rho} + \eta ^{\mu \rho} \eta ^{\nu \sigma}  -\eta ^{\mu \nu} \eta ^{\rho \sigma} 
\end{align*}
Using this form, we start by calculating the double functional derivative of $I_{1 2}$ multiplied with $x_{\nu}(s_1)$. Using $(\ref{rule1})$ we find:
\begin{align*}
&\quad x_{\nu}(s_1)\dfrac{\delta}{\delta x^{\rho}(s_1)} \dfrac{\delta}{\delta x^{\sigma}(s_2)} \frac{\dot{x}(\tau _1)  \dot{x}(\tau_2) - |\dot{x}(\tau_1)| |\dot{x}(\tau_2)|}{(x(\tau_1) -x(\tau_2))^2} = \\
&= \frac{x_{\nu}(s_1)}{\left(x_1-x_2\right)^2} \bigg[ \left( \eta _{\rho \sigma} - \frac{\dot{x}_{\sigma}(\tau_2)\dot{x}_{\rho}(\tau_1)}{|\dot{x}(\tau_2)||\dot{x}(\tau_1)|} \right) \partial _{\tau_1} \delta(\tau_1 - s_1) \partial _{\tau_2} \delta(\tau_2 - s_2)\\
& \quad + \left( \eta _{\rho \sigma} - \frac{\dot{x}_{\sigma}(\tau_1)\dot{x}_{\rho}(\tau_2)}{|\dot{x}(\tau_1)||\dot{x}(\tau_2)|} \right) \partial _{\tau_2} \delta(\tau_2 - s_1) \partial _{\tau_1} \delta(\tau_1 - s_2) \\
& \quad + \left( \frac{\dot{x}_{\sigma}(\tau_1)\dot{x}_{\rho}(\tau_1)|\dot{x}(\tau_2)|}{|\dot{x}(\tau_1)|^3} - \eta _{\rho \sigma} \frac{|\dot{x}(\tau_2)|}{|\dot{x}(\tau_1)|} \right) \partial _{\tau_1} \delta(\tau_1 - s_1) \partial _{\tau_1} \delta(\tau_1 - s_2) \\
& \quad + \left( \frac{\dot{x}_{\sigma}(\tau_2)\dot{x}_{\rho}(\tau_2)|\dot{x}(\tau_1)|}{|\dot{x}(\tau_2)|^3} - \eta _{\rho \sigma} \frac{|\dot{x}(\tau_1)|}{|\dot{x}(\tau_2)|} \right) \partial _{\tau_2} \delta(\tau_2 - s_1) \partial _{\tau_2} \delta(\tau_2 - s_2) \bigg]\\
& \quad - \frac{2 x_{\nu}(s_1)}{(x_1-x_2)^4} \bigg\{ ( x_{\rho}(\tau_1) -  x_{\rho}(\tau_2) ) ( \delta(\tau_1 - s_1) - \delta(\tau_2 - s_1)) \bigg[ \dot{x}_{\sigma}(\tau_1) \partial_{\tau_2} \delta (\tau_2 - s_2) \\
& \quad + \dot{x}_{\sigma}(\tau_2) \partial_{\tau_1} \delta (\tau_1 - s_2) -  \frac{|\dot{x}(\tau_1)|}{|\dot{x}(\tau_2)|}\dot{x}_{\sigma}(\tau_2)\partial_{\tau_2} \delta (\tau_2 - s_2) - \frac{|\dot{x}(\tau_2)|}{|\dot{x}(\tau_1)|}\dot{x}_{\sigma}(\tau_1)\partial_{\tau_1} \delta (\tau_1 - s_2) \bigg] \\
& \quad + ( x_{\sigma}(\tau_1) -  x_{\sigma}(\tau_2) ) ( \delta(\tau_1 - s_2) - \delta(\tau_2 - s_2)) \bigg[ \dot{x}_{\rho}(\tau_1) \partial_{\tau_2} \delta (\tau_2 - s_1) \\
& \quad + \dot{x}_{\rho}(\tau_2) \partial_{\tau_1} \delta (\tau_1 - s_1) -  \frac{|\dot{x}(\tau_1)|}{|\dot{x}(\tau_2)|}\dot{x}_{\rho}(\tau_2)\partial_{\tau_2} \delta (\tau_2 - s_1) - \frac{|\dot{x}(\tau_2)|}{|\dot{x}(\tau_1)|}\dot{x}_{\rho}(\tau_1)\partial_{\tau_1} \delta (\tau_1 - s_1) \bigg] \\
& \quad + ( \dot{x_1}  \dot{x_2} - |\dot{x_1}| |\dot{x_2}| ) \eta_{\rho \sigma} ( \delta(\tau_1 - s_2) - \delta(\tau_2 - s_2))( \delta(\tau_1 - s_1) - \delta(\tau_2 - s_1)) \bigg\} \\
& \quad + \frac{8 x_{\nu}(s_1)}{(x_1-x_2)^6} ( \dot{x_1}  \dot{x_2} - |\dot{x_1}| |\dot{x_2}| ) ( x_{\sigma}(\tau_1) -  x_{\sigma}(\tau_2)) ( x_{\rho}(\tau_1) -  x_{\rho}(\tau_2)) \cdot \\
& \quad \cdot \:( \delta(\tau_1 - s_2) - \delta(\tau_2 - s_2))( \delta(\tau_1 - s_1) - \delta(\tau_2 - s_1)) \, .
\end{align*}
We order the above result by the structure of the delta functions and derivatives that appear in it. To abbreviate, we use that the Wilson loop integral is symmetric under $(\tau_1 \leftrightarrow \tau_2)$. Also, we fix the parametrization to be of unit-speed, demanding that $|\dot{x}|= i$. Then we get the following expression (where by $\hat{=}$ we mean that the expression on the right-hand side gives the same result when integrated over $\tau_1$ and $\tau_2$ in parametrization by arc-length):
\begin{align}
&\quad x_{\nu}(s_1)\dfrac{\delta}{\delta x^{\rho}(s_1)} \dfrac{\delta}{\delta x^{\sigma}(s_2)} \frac{\dot{x}(\tau _1)  \dot{x}(\tau_2) - |\dot{x}(\tau_1)| |\dot{x}(\tau_2)|}{(x(\tau_1) -x(\tau_2))^2} \quad \hat{=} \notag \\
&\hat{=} \: x_{\nu}(s_1) \big\{ F^{(1)}_{\rho \sigma}(\tau_1,\tau_2)  \partial _{\tau_1} \delta(\tau_1 - s_1) \partial _{\tau_1} \delta(\tau_1 - s_2) + F^{(2)}_{\rho \sigma}(\tau_1,\tau_2) \partial _{\tau_1} \delta(\tau_1 - s_1) \partial _{\tau_2} \delta(\tau_2 - s_2) \notag \\
& \quad + F^{(3)}_{\rho \sigma}(\tau_1,\tau_2) (\delta(\tau_1 - s_1)  - \delta(\tau_2 - s_1)) \partial_{\tau_1} \delta(\tau_1 - s_2) \notag \\
& \quad + F^{(3)}_{\sigma \rho}(\tau_1,\tau_2) (\delta(\tau_1 - s_2)  - \delta(\tau_2 - s_2)) \partial_{\tau_1} \delta(\tau_1 - s_1) \notag \\
& \quad + F^{(4)}_{\rho \sigma}(\tau_1,\tau_2) (\delta(\tau_1 - s_2)  - \delta(\tau_2 - s_2)) (\delta(\tau_1 - s_1)  - \delta(\tau_2 - s_1)) \big\} \label{deriv}
\end{align}
Here, we defined:
\begin{align}
F^{(1)}_{\rho \sigma}(\tau_1,\tau_2) :&= - \frac{2}{x_{12}^2}  \left( \dot{x}_{1 \rho} \dot{x}_{1 \sigma} + \eta_{\rho \sigma}     \right) \, , \qquad
F^{(2)}_{\rho \sigma}(\tau_1,\tau_2) := \frac{2}{x_{12}^2}  \left( \dot{x}_{1 \rho} \dot{x}_{2 \sigma} + \eta_{\rho \sigma}     \right) \label{F1} \, ,\\
F^{(3)}_{\rho \sigma}(\tau_1,\tau_2) :&= \frac{4}{x_{12}^4}  \, x_{12 \rho} \, \dot{x}_{12 \sigma} \, , \qquad
F^{(4)}_{\rho \sigma}(\tau_1,\tau_2) := 2 \: \frac{ \dot{x_1}  \dot{x_2} + 1}{x_{12}^4} \left( 4 \, \frac{x_{12 \rho} x_{12 \sigma}}{x_{12}^2}  - \eta _{\rho \sigma} \right) \, . \label{F2}
\end{align}
We will denote the contribution to $P^{(1) \, \mu} _{\text{bos}, \, \varepsilon} \vev{W(C)}_{(1)} $ of the above terms by $C_i^{\mu}$, i.e.:
\begin{align}
C_i^{\mu} & := \int \limits _{0} ^L \diff \tau_1 \diff \tau_2 \diff s_1 \diff s_2 \big\{ \left( \theta(s_2-s_1- \varepsilon) c_i^{\mu}(\tau_1 , \tau_2 , s_1, s_2) \right) + \left( \varepsilon \rightarrow - \varepsilon \right) \big\} 
\end{align}
and we define:
\begin{align}
c_1^{\mu}(\tau_1 , \tau_2 , s_1, s_2)  & := N^{\mu \nu \rho \sigma} x_{\nu}(s_1)F^{(1)}_{\rho \sigma}(\tau_1,\tau_2) \partial _{\tau_1} \delta(\tau_1 - s_1) \partial _{\tau_1} \delta(\tau_1 - s_2)\\
c_2^{\mu}(\tau_1 , \tau_2 , s_1, s_2)  & := N^{\mu \nu \rho \sigma} x_{\nu}(s_1) F^{(2)}_{\rho \sigma}(\tau_1,\tau_2)  \partial _{\tau_1} \delta(\tau_1 - s_1) \partial _{\tau_2} \delta(\tau_2 - s_2) \\
c_3^{\mu}(\tau_1 , \tau_2 , s_1, s_2)  &:= N^{\mu \nu \rho \sigma} x_{\nu}(s_1) F^{(3)}_{\rho \sigma}(\tau_1,\tau_2)(\delta(\tau_1 - s_1) - \delta(\tau_2 - s_1) ) \partial_{\tau_1} \delta(\tau_1 - s_2) \\
c_4^{\mu}(\tau_1 , \tau_2 , s_1, s_2)  &:= N^{\mu \nu \rho \sigma} x_{\nu}(s_1) F^{(3)}_{\sigma \rho}(\tau_1,\tau_2) ( \delta(\tau_1 - s_2) - \delta(\tau_2 - s_2)) \partial_{\tau_1} \delta(\tau_1 - s_1) \\
c_5^{\mu}(\tau_1 , \tau_2 , s_1, s_2)  &:= N^{\mu \nu \rho \sigma} x_{\nu}(s_1) F^{(5)}_{\rho \sigma}(\tau_1,\tau_2) (\delta(\tau_1 - s_2)  - \delta(\tau_2 - s_2)) \cdot \notag \\ 
& \quad \cdot (\delta(\tau_1 - s_1)  - \delta(\tau_2 - s_1))
\end{align} 
With these definitions, we have:
\begin{align*}
P^{(1) \, \mu} _{\text{bos}, \, \varepsilon} \vev{W(C)} _{(1)} =  - \frac{\lambda }{16 \pi^2}  \sum \limits _{i=1}^{5} C_i^{\mu} \mbox{ ,}
\end{align*} 
We first discuss these terms separately, integrating out the delta-functions. For explicitness, we spell out the calculation for $C_1^{\mu}$. Consider the integral
\begin{align*}
& \quad \int _0 ^L \diff \tau_1 \diff \tau_2 F^{(1)}_{\rho \sigma}(\tau_1,\tau_2) \int _0 ^L \diff s_1 \diff s_2 \theta (s_2-s_1-\varepsilon) x_{\nu}(s_1) \partial _{\tau_1} \delta(\tau_1 - s_1) \partial _{\tau_1} \delta(\tau_1 - s_2) \\
&= - \int _0 ^L \diff \tau_1 \diff \tau_2 F^{(1)}_{\rho \sigma}(\tau_1,\tau_2) \int _0 ^L \diff s_1 x_{\nu}(s_1) \partial _{s_1} \delta(\tau_1 - s_1) \delta(\tau_1 - s_1 - \varepsilon) \\
&=  \int _0 ^L \diff \tau_1 \diff \tau_2 F^{(1)}_{\rho \sigma}(\tau_1,\tau_2) \int _0 ^L \diff s_1 \delta(\tau_1 - s_1)\partial _{s_1} ( x_{\nu}(s_1) \delta(\tau_1 - s_1 - \varepsilon)) \\
&=  \int _0 ^L \diff \tau_1 \diff \tau_2 F^{(1)}_{\rho \sigma}(\tau_1,\tau_2)   \dot{x}_{\nu}(\tau_1) \delta(-\varepsilon) \\*
& \quad + \int _0 ^L \diff \tau_1 \diff \tau_2 F^{(1)}_{\rho \sigma}(\tau_1,\tau_2) \int _0 ^L \diff s_1 x_{\nu}(s_1) \delta(\tau_1 - s_1)\partial _{\varepsilon}  \delta(\tau_1 - s_1 - \varepsilon) \\
&= \int _0 ^L \diff \tau_1 \diff \tau_2 F^{(1)}_{\rho \sigma}(\tau_1,\tau_2)   \dot{x}_{\nu}(\tau_1) \delta(\varepsilon) + \partial _{\varepsilon} \int _0 ^L \diff \tau_1 \diff \tau_2 F^{(1)}_{\rho \sigma}(\tau_1,\tau_2)   x_{\nu}(\tau_1) \delta(\varepsilon)
\end{align*}
Using that $\delta(\varepsilon) = \delta(-\varepsilon) $ and $ \partial_{\varepsilon}  \delta(\varepsilon) = - \partial_{- \varepsilon} \delta(-\varepsilon) $ we get:
\begin{align}
C_1^{\mu} = 2 \int _0 ^L \diff \tau_1 \diff \tau_2 N^{\mu \nu \rho \sigma}   F^{(1)}_{\rho \sigma}(\tau_1,\tau_2)   \dot{x}_{\nu}(\tau_1) \delta(\varepsilon) \label{C1}
\end{align}
Using the periodicity of the curves we consider, one may convince oneself that all boundary terms in the above calculation vanish. In a similar fashion we get the following results:
\begin{align}
C_2^{\mu} &= \int _0 ^L \diff \tau_1 \diff \tau_2 N^{\mu \nu \rho \sigma} F^{(2)}_{\rho \sigma}(\tau_1,\tau_2) \dot{x}_{\nu}(\tau_1) \left( \delta(\tau_2 - \tau_1 -\varepsilon)+\delta(\tau_2 - \tau_1 +\varepsilon) \right) \notag \\
& + \partial _{\varepsilon} \int _0 ^L \diff \tau_1 \diff \tau_2 N^{\mu \nu \rho \sigma} F^{(2)}_{\rho \sigma}(\tau_1,\tau_2)  x_{\nu}(\tau_1)  \left( \delta(\tau_2 - \tau_1 -\varepsilon)- \delta(\tau_2 - \tau_1 +\varepsilon) \right) \label{C2}  \\
C_3^{\mu} &= 2 \int _0 ^L \diff \tau_1 \diff \tau_2 N^{\mu \nu \rho \sigma} F^{(3)}_{\rho \sigma}(\tau_1,\tau_2)  x_{\nu}(\tau_1)   \delta(\varepsilon) \notag \\
&  \quad - \int _0 ^L \diff \tau_1 \diff \tau_2 N^{\mu \nu \rho \sigma} F^{(3)}_{\rho \sigma}(\tau_1,\tau_2)  x_{\nu}(\tau_1)\left( \delta(\tau_1 - \tau_2 -\varepsilon) + \delta(\tau_1 - \tau_2 +\varepsilon) \right) \label{C3} \\
C_4^{\mu} & =  \int _0 ^L \diff \tau_1 \diff \tau_2 N^{\mu \nu \rho \sigma} F^{(3)}_{\sigma \rho}(\tau_1,\tau_2) \dot{x}_{\nu}(\tau_1) - 2 \int _0 ^L \diff \tau_1 \diff \tau_2 N^{\mu \nu \rho \sigma}F^{(3)}_{\sigma \rho}(\tau_1,\tau_2) x_{\nu}(\tau_1) \delta (\varepsilon)  \notag \\
 & \quad + \int _0 ^L\diff \tau_1 \diff \tau_2 N^{\mu \nu \rho \sigma} F^{(3)}_{\sigma \rho}(\tau_1,\tau_2)   x_{\nu}(\tau_1) ( \delta (\tau_2 -\tau_1 -\varepsilon) + \delta (\tau_2 -\tau_1 +\varepsilon)) \notag \\
& \quad -  \int _0 ^L \diff \tau_1 \diff \tau_2 N^{\mu \nu \rho \sigma} F^{(3)}_{\sigma \rho}(\tau_1,\tau_2)     \dot{x}_{\nu}(\tau_1) \left( \theta (\tau_2 -\tau_1 -\varepsilon) + \theta (\tau_2 -\tau_1 +\varepsilon) \right) \label{C4} \\
C_5^{\mu} &= - 2 \int _0 ^L\diff \tau_1 \diff \tau_2 N^{\mu \nu \rho \sigma} F^{(4)}_{\rho \sigma}(\tau_1,\tau_2)\left( x_{\nu}(\tau_1) - x_{\nu}(\tau_2) \right) \theta(\tau_2 - \tau_1 -\varepsilon) \label{C5}
\end{align}
Taking into account that $N^{\mu \nu \rho \sigma} A_{[\rho \sigma] \nu} = 0$, we can simplify $C_3^{\mu} + C_4^{\mu}$ to get:
\begin{align}
C_3^{\mu} + C_4^{\mu} & = \int _0 ^L\diff \tau_1 \diff \tau_2 N^{\mu \nu \rho \sigma} F^{(3)}_{\sigma \rho}(\tau_1,\tau_2) \dot{x}_{\nu}(\tau_1) \notag \\
& \quad -  \int _0 ^L \diff \tau_1 \diff \tau_2 N^{\mu \nu \rho \sigma} F^{(3)}_{\sigma \rho}(\tau_1,\tau_2)     \dot{x}_{\nu}(\tau_1) \left( \theta (\tau_2 -\tau_1 -\varepsilon) + \theta (\tau_2 -\tau_1 +\varepsilon) \right) \notag \\
&= - \int _0 ^L \diff \tau_1 \diff \tau_2 N^{\mu \nu \rho \sigma} F^{(3)}_{\sigma \rho}(\tau_1,\tau_2) (\dot{x}_{\nu}(\tau_1) - \dot{x}_{\nu}(\tau_2)) \left( \theta (\tau_2 -\tau_1 -\varepsilon) \right) \label{C34} 
\end{align}
Inserting the contractions
\begin{align}
	N^{\mu \nu \rho \sigma}   F^{(1)}_{\rho \sigma}(\tau_1,\tau_2)   \dot{x}_{1 \nu} &=  6 \, \frac{ \dot{x}_1 ^{\mu}}{x_{12}^2 } \quad ,
	\quad  N^{\mu \nu \rho \sigma}  F^{(2)}_{\rho \sigma}(\tau_1,\tau_2)   \dot{x}_{1 \nu} = -2 \, \frac{2\dot{x}_1 ^{\mu} + \dot{x}_2 ^{\mu}}{x_{12}^2} \, , \\
	N^{\mu \nu \rho \sigma}   F^{(2)}_{\rho \sigma}(\tau_1,\tau_2)   x_{1 \nu} &= -\frac{2}{x_{12}^2} \left( 2 x_1 ^{\mu} + x_1^{\mu} (\dot{x}_1 \dot{x}_2) - \dot{x}_1^{\mu} (x_1 \dot{x}_2) - \dot{x}_2^{\mu} (x_1 \dot{x}_1)  \right) \, \\
	N^{\mu \nu \rho \sigma}   F^{(3)}_{\sigma \rho }(\tau_1,\tau_2)   \dot{x}_{12 \nu} &= -8 \, \frac{\dot{x}_1 \dot{x}_2 + 1 }{x_{12}^4} \, x_{12}^{\mu}  \quad ,
	\quad  N^{\mu \nu \rho \sigma}   F^{(4)}_{\rho \sigma}(\tau_1,\tau_2)   x_{12 \nu}  = 12 \, \frac{\dot{x}_1 \dot{x}_2 + 1 }{x_{12}^4} x_{12}^{\mu}
\end{align}
into $ (\mathrm{\ref{C1}}) - (\mathrm{\ref{C34}})$ and using the $(\tau_1 \leftrightarrow \tau_2)$-symmetry of the integral we arrive at:
\begin{align}
& P^{(1) \, \mu} _{\text{bos}, \, \varepsilon}\vev{W(C)}_{(1)}  =  - \frac{\lambda }{16 \pi^2} \sum \limits _{i=1} ^{5} C_i^{\mu}  =   \frac{\lambda }{16 \pi^2} \bigg\lbrace  -6 \int _0 ^L \diff \tau_1 \diff \tau_2 \frac{\dx_1^{\mu} + \dx_2^{\mu}  }{(x_1-x_2)^2}  \, \delta(\varepsilon) \notag \\
& + 6 \int _0 ^L \diff \tau_1 \diff \tau_2 \frac{\dx_1^{\mu}}{(x_1-x_2)^2}  (\delta(\tau_2 - \tau_1 - \varepsilon) + \delta(\tau_2 - \tau_1 + \varepsilon)) \notag \\
& + 2 \, \partial_{\varepsilon} \int _0 ^L \diff \tau_1 \diff \tau_2 \frac{2 \, x_1^{\mu} + x_1^{\mu} (\dx_1 \dx_2) - \dx_1^{\mu} (x_1 \dx_2) - \dx_2^{\mu} (x_1 \dx_1) }{(x_1-x_2)^2} \, \big( \delta(\tau_2 - \tau_1 - \varepsilon) - \delta(\tau_2 - \tau_1 + \varepsilon) \big) \notag \\
&  + 16 \int _0 ^L \diff \tau_1 \diff \tau_2 \frac{\dx_1  \dx_2 +1}{(x_1-x_2)^4}\:(x_1-x_2)^{\mu} \theta ( \tau_2 - \tau_1 - \varepsilon) \bigg\rbrace \label{res1}
\end{align}
We now expand the two middle terms. Making use of
\begin{align*}
\dot{x}^2 \equiv - 1 \Rightarrow \dot{x}\ddot{x} \equiv 0 \Rightarrow \ddot{x}^2 + \dot{x} \dddot{x} \equiv 0 
\end{align*}
we find:
\begin{align}
6 \int _0 ^L \diff \tau_1 \diff \tau_2 \frac{\dot{x}_1^{\mu}}{(x_1-x_2)^2}  (\delta(\tau_2 - \tau_1 - \varepsilon) + \delta(\tau_2 - \tau_1 + \varepsilon)) = \notag \\
= - \frac{12}{\varepsilon ^2} \int \diff \tau \dot{x}^{\mu}(\tau) +  \int _0 ^L \diff \tau \dot{x}^{\mu}(\tau) \ddot{x}(\tau)^2 + \mathcal{O}(\varepsilon) \label{expan1}
\end{align} 
and
\begin{align}
&2 \partial_{\varepsilon} \int _0 ^L \diff \tau_1 \diff \tau_2 \frac{2 x_1^{\mu} + x_1^{\mu} (\dot{x}_1 \dot{x}_2) - \dot{x}_1^{\mu} (x_1 \dot{x}_2) - \dot{x}_2^{\mu} (x_1 \dot{x}_1) }{(x_1-x_2)^2} (\delta(\tau_2 - \tau_1 - \varepsilon) - \delta(\tau_2 - \tau_1 + \varepsilon)) = \notag \\
&= - \: \frac{2}{\varepsilon^2} \int \diff \tau \dot{x}^{\mu} (\tau) + \frac{4}{3} \int _0 ^L \diff \tau \: \dddot{x}^\mu (\tau) - \frac{5}{6} \int _0 ^L \diff \tau \dot{x}^{\mu} (\tau) \ddot{x}^2(\tau) + \mathcal{O} (\varepsilon)   \label{expan2}
\end{align} 
Inserting $(\ref{expan1})$ and $(\ref{expan2})$ into $(\ref{res1})$ and using that 
\begin{align*}
\int \diff \tau \dot{x}^{\mu}(\tau) = 0 = \int _0 ^L \diff \tau \: \dddot{x}^\mu (\tau) \, ,
\end{align*}
which certainly holds for smooth periodic curves, we get:
\begin{align}
P^{(1) \, \mu} _{\text{bos}, \, \varepsilon} \vev{W(C)}_{(1)}  & =   \frac{\lambda}{16 \pi^2 } \bigg\lbrace \frac{1}{6} \int _0 ^L \diff \tau \: \dot{x}^{\mu} (\tau) \ddot{x}^2(\tau)
 -6 \int _0 ^L \diff \tau_1 \diff \tau_2 \frac{\dx_1^{\mu} + \dx_2^{\mu}  }{(x_1-x_2)^2}  \, \delta(\varepsilon)  \notag  \\
& \quad + 16 \int _0 ^L  \diff \tau_1 \diff \tau_2 \frac{\dot{x}_1  \dot{x}_2 + 1}{(x_1-x_2)^4}\:(x_1-x_2)^{\mu} \theta ( \tau_2 - \tau_1 - \varepsilon) + \mathcal{O} (\varepsilon)  \bigg\rbrace  \label{res2}
\end{align}
For any finite value of $\varepsilon$ we have $\delta(\varepsilon)=0$ and we therefore drop
the $\delta(\epsilon)$ term. Moreover
in the above result, the parametrization of the curve is still fixed to arc-length.  Lifting this constraint we have the reparametrization invariant result
\begin{align}
P^{(1) \, \mu} _{\text{bos}, \, \varepsilon} \vev{W(C)}_{(1)} & =  \frac{\lambda}{16 \pi^2 } \bigg\lbrace \frac{1}{6} \int \diff \tau \: \dot{x}^{\mu} (\tau) \left ( \frac{\ddot{x}^{2}}{\dot{x}^{4}} - \frac{(\dot{x}\cdot\ddot{x})^{2}}{\dot{x}^{6}} \, \right )+ \mathcal{O} (\varepsilon) \notag  \\
& \quad + 16 \int  \diff \tau_1 \diff \tau_2 \frac{\dot{x}_1  \dot{x}_2 - |\dot{x}_1| |\dot{x}_2| }{(x_1-x_2)^4}\:(x_1-x_2)^{\mu} \theta ( \tau_2 - \tau_1 - d(\tau_2,\varepsilon))  \bigg\rbrace \label{eqn:bosonic_result}
\end{align}

\subsection{Supersymmetric completion of the Maldacena-Wilson loop}
\label{app: Susycompl}

As we argued in section \ref{sec: supsymcomp} it will be enough for our purpose to require that the equations \begin{align}
\alg{Q}^{\a}_{A} (I) = Q^{\a}_{A} (I)\, \qquad
\bar{\alg{Q}}^{A\, \da} (I) =  {\bar Q}^{A\, \da}(I) 
\label{eqn: fvareqldiff}
\end{align}
with the supersymmetrically completed exponent
\begin{align}
I \, [A,\psi,\bar\psi,\phi; 
x,\theta,\btheta]=\oint _C \diff \tau \, \left( \mathcal{I}_{0} + \mathcal{I}_{1}+ \bar{\mathcal{I}}_{1} + \mathcal{I}_{2m} + \mathcal{I}_{2} + \bar{\mathcal{I}}_{2} +\cO({\bar\theta}^{i}{\theta}^{3-i}) \right)
\end{align}
hold true up to order $\theta$ and $\btheta$, respectively. Furthermore, we will only focus on the terms which are linear in the fields since only those will contribute to the 1-loop expectation value $\left\langle \mathcal{W}(C) \right\rangle_{(1)}$. Therefore, the equations \eqref{eqn: fvareqldiff} can be split into the following set of equations: 
\begin{align}
\label{eqn: QQ1}
\alg{Q}^{\a}_{A} (I_0) &=  \left. Q^{\a}_{A} (I_1) \right|_{\btheta=0} \\
\label{eqn: QQ2}
\alg{Q}^{\a}_{A} (I_1) &=  \left. Q^{\a}_{A} (I_2) \right|_{\btheta=0} \\
\label{eqn: QQ3}
\alg{Q}^{\a}_{A} (\bar{I}_1) &=  \left. Q^{\a}_{A} (I_0+I_{2m}) \right|_{\theta=\dot\theta=0} \\
\label{eqn: QQ4}
\bar{\alg{Q}}^{A\, \da} (I_0) &=  \left. {\bar Q}^{A\, \da} (\bar{I}_1) \right|_{\theta=0} \\
\label{eqn: QQ5}
\bar{\alg{Q}}^{A\, \da} (\bar{I}_1) &=  \left. {\bar Q}^{A\, \da} (\bar{I}_2) \right|_{\theta=0} \\
\label{eqn: QQ6}
\bar{\alg{Q}}^{A\, \da} (I_1) &=  \left. {\bar Q}^{A\, \da} (I_0+I_{2m}) \right|_{\btheta=\dot\btheta=0} 
\end{align}
where we introduced the notation
\begin{align}
I_x = \oint_C \diff \tau \, \mathcal{I}_x \, .
\end{align}
Let us start by calculating how $\mathcal{I}_0$ transforms under supersymmetry transformations generated by $\alg{Q}^{\a}_{A}$ and $\bar{\alg{Q}}^{A\, \da}$, respectively. In order to have a more compact notation we will mostly consider the equations \eqref{eqn: QQ1}-\eqref{eqn: QQ6} on the level of the integrand and only write the integral if we use integration by parts. With the basic field transformations given by \eqref{eqn: ftrans1}-\eqref{eqn: ftrans2} one finds 
\begin{align}
\alg{Q}^{\a}_{A} (\mathcal{I}_0) &=  i \epsilon^{\a \b} \, \bpsi^{\db}_{A} \, \dx_{\b \db} -  \sqrt{2} \, i  \, \psi^{D \a} \, \bar\eta_{AD} \, |\dx|  \\
\bar{\alg{Q}}^{A\, \da} (\mathcal{I}_0) &= -  i \, \epsilon^{\da \db} \, \psi^{A \b} \, \dx_{\b \db}  -  \sqrt{2} \, i \, \bpsi^{\da}_{D} \, \eta^{AD} \, |\dx| \, . 
\end{align} 
It can easily be seen that the equations \eqref{eqn: QQ1} and \eqref{eqn: QQ4} are satisfied if we choose $\mathcal{I}_1$ and $\bar{\mathcal{I}}_1$ as follows:
\begin{align}
\mathcal{I}_{1} & = i \, \theta^{B \b} \, \bpsi^{\db}_{B} \, \dx_{\b \db} +  \sqrt{2} \, i \,
\theta^{C}_{\b} \, \psi^{D \b} \, \bar\eta_{CD} \, |\dx|   \\
\bar{\mathcal{I}}_{1} & =  -  i \, \btheta_{B}^{\db} \, \psi^{B \b} \, \dx_{\b \db}  -  \sqrt{2} \, i \, \btheta_{C \db} \, \bpsi^{\db}_{D} \, \eta^{CD} \, |\dx|    
\end{align}
Since we know how $\alg{Q}^{\a}_{A}$ and $\bar{\alg{Q}}^{A\, \da}$ act on fields we can directly write down how $\mathcal{I}_{1}$ and $\bar{ \mathcal{I}}_{1}$ transform under supersymmetry transformations.
\begin{align}
\alg{Q}^{\a}_{A}(\mathcal{I}_{1}) &= \sqrt{2} \, i  \, \theta^{B \b}  \Bigl(\partial^{\db \a}    \, \bar{\phi}_{AB} \Bigr)  \dx_{\b \db} + \ft{1}{\sqrt{2}} \, \theta^{C}_{\b} \, F^{\a \b}_{lin} \, \bar{\eta}_{CA} \, |\dx| \\
\alg{Q}^{\a}_{A}(\bar{\mathcal{I}}_{1}) &= -\ft{1}{2} \, \btheta_{A}^{\db} F^{\a \b}_{lin} \dx_{\b \db} - 2 \, i  \, \btheta_{C \db}  \Bigl(\partial^{\db \a}    \, \bar{\phi}_{AB} \Bigr) \eta^{CB}\, |\dx| \\
 \bar{\alg{Q}}^{A\, \da}(\mathcal{I}_{1}) &= -\ft{1}{2} \, \theta^{A \b} F^{\da \db}_{lin} \dx_{\b \db} + 2 \, i  \, \theta^{C}_{\b}  \Bigl(\partial^{\b \da}    \, \phi^{AB} \Bigr) \bar\eta_{CB}\, |\dx|   \\    
\bar{\alg{Q}}^{A\, \da}(\bar{\mathcal{I}}_{1}) &= - \sqrt{2} \, i  \, \btheta_{B}^{\db}  \Bigl(\partial^{\b \da}    \, \phi^{AB} \Bigr)  \dx_{\b \db} + \ft{1}{\sqrt{2}} \, \btheta_{C \db} \, F^{\da \db}_{lin} \, \eta^{CA}\, |\dx|  
\end{align}
In these equations $F^{\a \b}_{lin}$ and $F^{\da \db}_{lin}$ denote the parts of \eqref{eqn: Fmunubispin} which are linear in the gauge fields. The parts $\mathcal{I}_2$ and $\bar{\mathcal{I}}_2$ can be constructed by imposing that the equations \eqref{eqn: QQ2} and \eqref{eqn: QQ5} hold true. The result reads:  
\begin{align}
\mathcal{I}_{2} & = - \ft{i}{\sqrt{2}} \, \theta^{C}_{\g} \, \theta^{B \b}  \Bigl(\partial^{\db \g}    \, \bar{\phi}_{CB} \Bigr)  \dx_{\b \db} + \ft{1}{2 \sqrt{2}} \, \theta^{C}_{\b} \, \theta^{D}_{\g} \, F^{\g \b} _{lin} \, \bar\eta_{CD}\, |\dx| +  \sqrt{2} \, i \, \theta^{C}_{\g} \, \dot{\theta}^{B \g} \, \bar{\phi}_{CB}  \\
\bar{\mathcal{I}}_{2} & = -\ft{i}{\sqrt{2}} \, \btheta_{C \dg} \, \btheta_{B}^{\db} \Bigl(\partial^{\b \dg}  \, \phi^{CB} \Bigr) \dx_{\b \db} - \ft{1}{2 \sqrt{2}} \, \btheta_{C \db} \, \btheta_{D \dg}\, F^{\dg \db}_{lin} \, \eta^{CD} \, |\dx| +  \sqrt{2} \, i \, \btheta_{C \dg} \, \dot{\btheta}_{B}^{\dg} \, \phi^{CB} \, 
\end{align}
Since the calculations which show that the equations \eqref{eqn: QQ2} and \eqref{eqn: QQ5} are indeed satisfied are a little bit more involved we will give some details on at least one of them. Applying the $\theta$-derivative of      $Q^{\a}_{A}$ to $I_2$ yields:
\begin{align}
\left. Q^{\a}_{A}(I_2) \right|_{\btheta=0}=&\int \diff \tau \, \left(- \sqrt{2} \, i \, \theta^{B}_{\b} \Bigl(\partial^{\db (\a}  \, \bar{\phi}_{AB} \Bigr) \dx \indices{^{\b)} _{\db}} + \ft{1}{\sqrt{2}} \,\theta^C_{\b} \, F^{\a \b}_{lin} \, \bar\eta_{CA} |\dx|- \sqrt{2} \, i \, \dot{\theta}^{B \a} \, \bar{\phi}_{AB} \right) \nn \\
=& \int \diff \tau \,  \left( \sqrt{2}  \, i \, \theta^{B \b} \Bigl(\partial^{\db \a}  \, \bar{\phi}_{AB} \Bigr) \dx_{\b \db} - \ft{i}{\sqrt{2}} \, \theta^{B \a} \Bigl(\partial^{\db \b}  \, \bar{\phi}_{AB} \Bigr) \dx_{\b \db} \right. \nn \\
&\qquad \quad \left. + \ft{1}{\sqrt{2}} \,\theta^C_{\b} \, F^{\a \b}_{lin} \, \bar\eta_{CA} |\dx|- \sqrt{2} \, i \, \dot{\theta}^{B \a} \, \bar{\phi}_{AB} \right) \nn \\
=& \int \diff \tau \, \left(  \sqrt{2} \, i  \, \theta^{B \b}  \Bigl(\partial^{\db \a}    \, \bar{\phi}_{AB} \Bigr)  \dx_{\b \db} + \ft{1}{\sqrt{2}} \, \theta^{C}_{\b} \, F^{\a \b}_{lin} \, \bar{\eta}_{CA} \, |\dx| \right) 
\end{align}
In order to get the second line we used the identity \eqref{eqn: weylindpartition}. We note that the last term in the second line can be rewritten as a derivative with respect to the curve parameter $\tau$ acting on $\bar{\phi}_{AB}$. Using integration by parts we see that the rewritten term cancels the $\dot\theta$-term. In a similar manner it can be shown that $\bar{I}_2$ satisfies equation \eqref{eqn: QQ5}. Let us now turn to the construction of $\mathcal{I}_{2m}$. While $\mathcal{I}_2$ and $\bar{\mathcal{I}}_2$ are not necessarily needed for our purpose since their contractions do not contribute to the bosonic order after having applied the level-one momentum generator, this does not apply to $\mathcal{I}_{2m}$. In contrast to the construction of $\mathcal{I}_1$, $\bar{\mathcal{I}}_1$, $\mathcal{I}_2$ and $\bar{\mathcal{I}}_2$ we do now have two equations for one expression and it is not clear that they are compatible with each other. We will start by calculating how $Q^{\a}_{A}$ acts on $I_0$.
\begin{align}
Q^{\a}_{A}(I_0)&=\int \diff s \, \diff \tau \, i \, \btheta_{A \da}(s)\, \dfrac{\delta}{\delta x_{\a\da}(s)} \, \Bigl(\ft{1}{2}  \, A^{\b \db} \, \dx_{\b \db} - \ft{1}{2} \, \phi^{CD} \, \bar\eta_{CD} \, |\dx|  \Bigr) \nn \\
&=\int \diff \tau  \left( \ft{i}{2} \, \btheta_{A \da}  \Bigl(\partial^{\a \da}  \, A^{\b \db} \Bigr) \dx_{\b \db} + i \, \dot{\btheta}_{A \da} \, A^{\a \da}  - \ft{i}{2} \, \btheta_{A \da} \, \Bigl( \partial^{\a \da} \,\phi^{CD} \Bigr) \bar\eta_{CD} \, |\dx|  - \ft{i}{2} \, \dot{\btheta}_{A \da} \, \phi^{CD} \,  \bar\eta_{CD} \ft{\dx^{\a \da}}{|\dx|} \right) \nn \\
&=\int \diff \tau \, \left( \ft{i}{2} \, \btheta_{A \da} \, \Bigl(\partial^{\a \da}  \, A^{\b \db} - \partial^{\b \db}  \, A^{\a \da} \Bigr) \dx_{\b \db} - \ft{i}{2} \, \btheta_{A \da} \, \Bigl( \partial^{\a \da} \,\phi^{CD} \Bigr) \bar\eta_{CD} \, |\dx|  - \ft{i}{2} \, \dot{\btheta}_{A \da} \, \phi^{CD} \,  \bar\eta_{CD} \ft{\dx^{\a \da}}{|\dx|} \right) \nn \\
&=\int \diff \tau \, \left( -\ft{1}{4} \, \btheta_{A}^{\db} \, F^{\a \b}_{lin} \, \dx_{\b \db} - \ft{1}{4} \, \btheta_{A \da} \, F^{\da \db}_{lin} \, \dx \indices{^{\a} _\db} -\ft{i}{2} \, \btheta_{A \da} \, \Bigl( \partial^{\a \da} \,\phi^{CD} \Bigr) \bar\eta_{CD} \, |\dx|  - \ft{i}{2} \, \dot{\btheta}_{A \da} \, \phi^{CD} \,  \bar\eta_{CD} \ft{\dx^{\a \da}}{|\dx|} \right) \nn
\end{align}
First we applied the functional derivative to $I_0$ and integrated out the $\delta$-functions by evaluating the generator integral. In order to get to the third line we used integration by parts in the second term. The last line follows by using the identity \eqref{eqn: F4decomp2}. The calculation including $\bar{Q}^{A \da}(I_0)$  works completely analogously.     
\begin{align}
\bar{Q}^{A \da}(I_0)&=\int \diff \tau \, \left( -\ft{1}{4} \, \theta^{A}_{\a} \, F^{\a \b}_{lin} \, \dx \indices{_{\b}^{\da}} - \ft{1}{4} \, \theta^{A \b} \, F^{\da \db}_{lin} \, \dx_{\b \db} +\ft{i}{2} \, \theta^A_{\a}  \Bigl( \partial^{\a \da} \,\phi^{CD} \Bigr) \bar\eta_{CD} \, |\dx|  + \ft{i}{2} \, \dot{\theta}^A_{\a} \, \phi^{CD} \,  \bar\eta_{CD} \ft{\dx^{\a \da}}{|\dx|} \right) \nn
\end{align}
By requiring that equation \eqref{eqn: QQ3} holds true, $\mathcal{I}_{2m}$ can be determined (up to the term including $\dot\theta$) to be
\begin{align}
\mathcal{I}_{2m} & = \ft{1}{4} \, \theta^{B}_{\g} \, \btheta_{B}^{\db} \, F^{\g \b}_{lin} \, \dx_{\b \db} + \ft{1}{4}\, \theta^{B \b} \, \btheta_{B \dg} \, F^{\dg \db}_{lin} \, \dx_{\b\db} + 2 \, i \, \theta^{B}_{\g} \, \btheta_{C \db}  \Bigl( \partial^{\db \g} \, \bar\phi_{B E} \Bigr)  \eta^{C E} \, |\dx| \nn \\ 
&   -\ft{i}{2} \, \theta^{B}_{\g} \, \btheta_{B \dg}  \Bigl(\partial ^{\g \dg} \, \phi^{C D} \Bigr) \bar\eta_{C D}  \, |\dx| + \ft{i}{2} \, \dot{\theta}^{B}_{\b}  \, \btheta_{B \db} \, \phi^{C D} \, \bar\eta_{C D} \, \ft{\dx^{\b \db}}{|\dx|}  - \ft{i}{2} \, \theta^{B}_{\b}  \, \dot{\btheta}_{B \db} \, \phi^{C D} \, \bar\eta_{C D}  \, \ft{\dx^{\b \db}}{|\dx|}  \, .
\label{eqn: I2m}
\end{align}
The application of $Q^{\a}_A$ to $\mathcal{I}_{2m}$ yields
\begin{align}
\left. Q^{\a}_A(\mathcal{I}_{2m}) \right|_{\theta=\dot\theta=0}=&-\ft{1}{4} \, \btheta_{A}^{\db} \, F^{\a \b}_{lin} \, \dx_{\b \db} + \ft{1}{4} \, \btheta_{A \da} \, F^{\da \db}_{lin} \, \dx \indices{^{\a} _\db} -2 \, i \, \btheta_{C \da}  \Bigl( \partial^{\a \da} \, \bar{\phi}_{AB} \Bigr) \eta^{CB} \, |\dx| \nn \\
&+ \ft{i}{2} \, \btheta_{A \da}  \Bigl( \partial^{\a \da} \,\phi^{CD} \Bigr) \bar\eta_{CD} \, |\dx|  + \ft{i}{2} \, \dot{\btheta}_{A \da} \, \phi^{CD} \,  \bar\eta_{CD} \ft{\dx^{\a \da}}{|\dx|} \, ,
\end{align}
from which we immediately see that equation \eqref{eqn: QQ3} is indeed satisfied. We will now show that \eqref{eqn: I2m} also solves equation \eqref{eqn: QQ6}. Therefore we calculate:
\begin{align}
\left. \bar{Q}^{A \da}(\mathcal{I}_{2m}) \right|_{\btheta=\dot\btheta=0}=&+\ft{1}{4} \, \theta^{A}_{\a} \, F^{\a \b}_{lin} \, \dx\indices{_\b ^\da} - \ft{1}{4} \, \theta^{A \a} \, F^{\da \db}_{lin} \, \dx \indices{_{\a \db}} -2 \, i \, \theta^B_\a  \Bigl( \partial^{\a \da} \, \bar{\phi}_{BC} \Bigr) \eta^{AC} \, |\dx| \nn \\
&+ \ft{i}{2} \, \theta^A_\a  \Bigl( \partial^{\a \da} \,\phi^{CD} \Bigr) \bar\eta_{CD} \, |\dx|  - \ft{i}{2} \, \dot{\theta}^A_\a \, \phi^{CD} \,  \bar\eta_{CD} \ft{\dx^{\a \da}}{|\dx|} 
\label{eqn: QbaronI2m}
\end{align}
The third term can be rewritten as follows
\begin{align}
2 \, i \, \theta^B_\a  \Bigl( \partial^{\a \da} \, \bar{\phi}_{BC} \Bigr) \eta^{AC} \, |\dx|=& \ft{i}{2} \, \theta^B_\a  \Bigl( \partial^{\a \da} \, \phi^{KL} \Bigr) \bar{\eta}_{MN} \, |\dx| \epsilon_{BCKL} \epsilon^{ACMN} \nn \\
=& i \, \theta^A_\a  \Bigl( \partial^{\a \da} \, \phi^{CD} \Bigr) \bar{\eta}_{CD} \, |\dx| 
-2 \, i \, \theta^C_\a  \Bigl( \partial^{\a \da} \, \phi^{AB} \Bigr) \bar{\eta}_{CB} \, |\dx| \, ,
\label{eqn: apleps6delta}
\end{align}
where we employed the identity \eqref{eqn: epseps6delta} . Inserting \eqref{eqn: apleps6delta} in \eqref{eqn: QbaronI2m} yields:
\begin{align}
\left. \bar{Q}^{A \da}(\mathcal{I}_{2m}) \right|_{\btheta=\dot\btheta=0}=&+\ft{1}{4} \, \theta^{A}_{\a} \, F^{\a \b}_{lin} \, \dx\indices{_\b ^\da} - \ft{1}{4} \, \theta^{A \b} \, F^{\da \db}_{lin} \, \dx \indices{_{\b \db}} - \ft{i}{2}\, \theta^A_\a  \Bigl( \partial^{\a \da} \, \phi^{CD} \Bigr) \bar{\eta}_{CD} \, |\dx| \nn \\
&+ 2 \, i \, \theta^C_\a  \Bigl( \partial^{\a \da} \, \phi^{AB} \Bigr) \bar{\eta}_{CB} \, |\dx|   - \ft{i}{2} \, \dot{\theta}^A_\a \, \phi^{CD} \,  \bar\eta_{CD} \ft{\dx^{\a \da}}{|\dx|} 
\label{eqn: QbaronI2mres}
\end{align}
If we combine this equation with the result for $\bar{Q}^{A \da}(I_0)$  we note that \eqref{eqn: QQ6} also holds true.

\subsection{Check of supersymmetry at one-loop order}
\label{app:qqbarsym}
The result (\ref{susyWilsonOneLoop}) should be supersymmetric by construction. Nevertheless it is a straightforward check to see if $Q^{\a}_{A}$ ($\bar Q^{A\,\da}$) annihilate $ \left\langle \mathcal{W}(C) \right\rangle_{(1)} $. We note that having computed the vacuum expectation value to order $\theta \btheta$ only allows us to check the invariance of $\left\langle \mathcal{W}(C) \right\rangle_{(1)} $ at order $\btheta$ ($\theta$) for $Q$ ($\bar{Q}$). Focusing on $Q$ for the moment, we verify that
\begin{align}
- \int \diff s \,   \frac{\delta}{\delta \theta^{A}_{\a}(s} \left\langle \mathcal{W}(C) \right\rangle_{(1)} 
+ i \int \diff s \, \btheta_{A\,\da}\, \frac{\delta}{\delta x_{\a\da}(s)} \Bigl( \left\langle \mathcal{W}(C) \right\rangle_{(1)}  \Bigr )_{\substack{\theta=0 \\ \btheta=0}} &= 0 \; . 
\end{align}
Combining the results of the individual terms
\begin{align}
- \int \diff s \,  \frac{\delta}{\delta \theta^{A}_{\a}(s)} \left\langle \mathcal{W}(C) \right\rangle_{(1)}  =&  \frac{i \lambda}{4 \pi^2}  \int \diff \tau_1 \diff \tau_2 \; \Bigl\{ \btheta_{A \da}(\tau_2) \, \frac{\dx_1  \dx_2 - \lvert \dx_1 \rvert \lvert \dx_2 \rvert}{(x_1-x_2)^4 }\:x_{12}^{\a \da}\nn \\
&+ \frac{\dot{\btheta}_{A \da}(\tau_2)}{2}  \frac{\dot{x}_1^{\a \da} }{x_{12}^2}
- \frac{\dot{\btheta}_{A \da}(\tau_2)}{2 x_{12}^2} \frac{\left| \dot{x}_1 \right|}{\left| \dot{x}_2 \right|}\dot{x}_2^{\a \da}\Bigr\} \nn \; ,
\end{align}
\begin{align}
i \int \diff s \, \btheta_{A\,\da}\, \frac{\delta}{\delta x_{\a\da}(s)} \Bigl( \left\langle \mathcal{W}(C) \right\rangle_{(1)}  \Bigr )_{\substack{\theta=0 \\ \btheta=0}} =& - \frac{i \lambda}{4 \pi^2}  \int \diff \tau_1 \diff \tau_2 \; \Bigl\{ \btheta_{A \da}(\tau_2)  \, \frac{\dx_1  \dx_2 - \lvert \dx_1 \rvert \lvert \dx_2 \rvert}{(x_1-x_2)^4 } \:x_{12}^{\a \da}\nn \\
&+ \frac{\dot{\btheta}_{A \da}(\tau_2)}{2} \frac{\dot{x}_1^{\a \da}}{x_{12}^2} 
 -  \frac{\dot{\btheta}_{A \da}(\tau_2)}{2 x_{12}^2}\frac{\left| \dot{x}_1 \right|}{\left| \dot{x}_2 \right|} \dot{x}_2^{\a \da}\Bigr \}
\end{align}
we find the expected result
\begin{align}
Q^{\a}_{A} \left\langle \mathcal{W}(C) \right\rangle_{(1)}  \Bigr|_{\btheta} = 0 \; .
\end{align}
The calculation for $\bar Q^{A\,\da} \left\langle \mathcal{W}(C) \right\rangle_{(1)}  \Bigr|_{\theta} = 0$ can be repeated equally.

\subsection{Non-local variation of the fermionic contributions to \texorpdfstring{$\vev{\mathcal{W}(C)}$}{<W(C)>}}
\label{app:Formulas}

According to \eqref{aa1}, the fermionic part of the generator $P^{(1) \, \mu} _{\varepsilon}$ is given by
\begin{align}
P^{(1) \, \mu} _{\text{ferm}, \, \varepsilon} 
= - \frac{i}{4} \int\limits_0^{L} \!\! ds_1\,  d s_2 \: \bar{q}^{A \da}(s_2) \bar{\sigma}^{\mu}_{\a \da} q_A^{\a}(s_1) \left( \theta(s_1 - s_2 - \epsilon) - \theta(s_2 - s_1 - \epsilon) \right) \, ,
\end{align}
with
\begin{align}
q^{\a}_{A}(s)  &=  - \dfrac{\delta}{\delta \theta^{A}_{\a}(s)}
+ i \, \btheta_{A \da}(s)\, \dfrac{\delta}{\delta x_{\a\da}(s)} \,  \\
{\bar q}^{A\, \da}(s)  &=   \dfrac{\delta}{\delta \btheta_{A \da}(s)}
-i \, \theta^{A}_{\a}(s)\, \dfrac{\delta}{\delta x_{\a\da}(s)} \,  \, .
\end{align}
We will only be interested in the fermionic correction to the bosonic result (\ref{eqn:bosonic_result}). Therefore we are looking for contributions where the final result does not depend on the Gra{\ss}mann variables $\theta$ and $\btheta$. This means that the only correction can come from the action of
\begin{align}
\hat{P}^{(1) \, \mu} _{\text{ferm}, \, \varepsilon} = \frac{i}{4} \int\limits_0^{L} \!\! ds_1\,  d s_2 \: \frac{\delta}{\delta \btheta_{A \da}(s_2)} \, \bar{\sigma}^{\mu}_{\a\da}\,  \frac{\delta}{\delta \theta^{A}_{\a}(s_1)} \left( \theta(s_1 - s_2 - \epsilon) - \theta(s_2 - s_1 - \epsilon) \right) \,
\end{align}
on objects that have a $\theta \btheta$ component.

To simplify the calculation it is useful to take a look at the $\theta \btheta$ structure of the one loop result (\ref{susyWilsonOneLoop}) and write down the action of the interesting part of the fermionic generator
\begin{align}
\label{eqn:QQb_on_local}
&\hat{P}^{(1) \, \mu} _{\text{ferm}, \, \varepsilon} \ \btheta_{B}(\tau_2)\bar{\sigma}_\nu \theta^B(\tau_1) = - i \,2  \delta^\mu_\nu \,
\Bigl [ \, \theta(\tau_{1}-\tau_{2}-\epsilon) -   \theta(\tau_{2}-\tau_{1}-\epsilon)\, \Bigr ] \\
&\hat{P}^{(1) \, \mu} _{\text{ferm}, \, \varepsilon} \ \btheta_{B}(\tau_2)\bar{\sigma}_\nu\dot{\theta}^B(\tau_1) = - i \,2  \delta^\mu_\nu \,
\Bigl [ \, \delta(\tau_{1}-\tau_{2}-\epsilon) +   \delta(\tau_{2}-\tau_{1}-\epsilon)\, \Bigr ]
\end{align}
and the corollaries
\begin{align}
&\hat{P}^{(1) \, \mu} _{\text{ferm}, \, \varepsilon} \ \btheta_{B}(\tau_1)\bar{\sigma}_\nu\theta^B(\tau_1) = 0 \\
&\hat{P}^{(1) \, \mu} _{\text{ferm}, \, \varepsilon} \ \btheta_{B}(\tau_2)\bar{\sigma}_{\nu}\dot{\theta}^B(\tau_1) = - \hat{P}^{(1) \, \mu} _{\text{ferm}, \, \varepsilon} \ \dot{\btheta}_{B}(\tau_2)\bar{\sigma}_{\nu}\theta^B(\tau_1) \\
\label{eqn:QQb_on_localdot}
&\hat{P}^{(1) \, \mu} _{\text{ferm}, \, \varepsilon} \left( \dot{\btheta}_{B}(\tau_1)\bar{\sigma}_\nu \theta^B(\tau_1) - \btheta_{B}(\tau_1)\bar{\sigma}_\nu \dot{\theta}^B(\tau_1)\right)  =  i \, 8 \; \delta(- \epsilon) \, \delta^\mu_\nu \; .
\end{align}
Using these relations we get
\begin{align}
& \hat{P}^{(1) \, \mu} _{\text{ferm}, \, \varepsilon} \left\langle \mathcal{W}(C) \right\rangle_{(1)} = \notag \\ 
& = - \frac{\lambda}{4 \pi^2}  \int _0 ^L \diff \tau_1  \diff \tau_2  \bigg \lbrace  \left(  \,2 \, \frac{\dot{x}_1 \dot{x}_2 +1}{x_{12}^4} \, x_{12}^{\mu}  + 2 i \, \frac{\epsilon^{\mu \nu \rho \kappa} \dot{x}_{1\,\nu} \dot{x}_{2\,\rho}x_{12\,\kappa}}{x_{12}^4}    \right)\, \Bigl [ \, \theta(\tau_{1}-\tau_{2}-\epsilon) -   \theta(\tau_{2}-\tau_{1}-\epsilon)\, \Bigr] \notag \\
& \qquad \qquad - \frac{\dot{x}_1^\mu + \dot{x}_2^\mu}{x_{12}^2}  \Bigl [ \, \delta(\tau_{1}-\tau_{2}-\epsilon) +   \delta(\tau_{2}-\tau_{1}-\epsilon)\, \Bigr ] -  4  \;  \frac{\dot{x}_2^\mu}{x_{12}^2} \delta(\epsilon) \bigg \rbrace \notag \\
& = - \frac{\lambda}{4 \pi^{2}}\, \bigg \lbrace  \int _0 ^L \diff \tau_1 \diff \tau_2  \left( 4 \, \frac{\dx_1  \dx_2 +1}{(x_1-x_2)^4}\: x_{12} ^{\mu} \, \theta ( \tau_2 - \tau_1 - \varepsilon)  - 2  \;  \frac{\dot{x}_1^\mu + \dot{x}_2^\mu}{x_{12}^2} \delta(\epsilon) \right) \notag \\
& \qquad \qquad + \frac{1}{3} \int _0 ^L \diff \tau \: \dot{x}^{\mu} (\tau) \ddot{x}^2(\tau)   + \mathcal{O} (\varepsilon) \bigg \rbrace
\label{here33}
\end{align}
where in the last step we replaced $(\tau_1 \leftrightarrow \tau_2)$ where appropriate and took the results of (\ref{expan1}) into account. Writing this again in a reparametrization invariant form and noting that $\delta(\varepsilon)=0$, we arrive at:
\begin{align}
\left. P^{(1) \, \mu} _{\text{ferm}, \, \varepsilon} \left\langle \mathcal{W}(C) \right\rangle_{(1)} \right|_{\substack{\theta=0 \\ \btheta=0}}  =
- \frac{\lambda}{16\pi^{2}}\, \bigg \lbrace & 16 \int \diff \tau_1 \diff \tau_2 \frac{\dx_1  \dx_2 - \lvert \dx_1 \rvert \lvert \dx_2 \rvert}{(x_1-x_2)^4}\:(x_1-x_2)^{\mu} \theta ( \tau_2 - \tau_1 - d(\tau_2,\varepsilon)) \nn\\
& + \frac{4}{3} \int \diff \tau \: \dot{x}^{\mu} (\tau) \left ( \frac{\ddot{x}^{2}}{\dot{x}^{4}} - \frac{(\dot{x}\cdot\ddot{x})^{2}}{\dot{x}^{6}} \, \right )+ \mathcal{O} (\varepsilon) \bigg \rbrace \, . 
\end{align}

\newpage

\bibliographystyle{nb}
\bibliography{botany}

\begin{thebibliography}{10}
\ifx\href\asklfhas\newcommand{\href}[2]{#2}\fi
\ifx\arxivref\asklfhas\newcommand{\arxivref}[2]{\href{http://arxiv.org/abs/#1}{#2}}\fi
\ifx\doiref\asklfhas\newcommand{\doiref}[2]{\href{http://dx.doi.org/#1}{#2}}\fi
\raggedright
\small
\parskip 0pt

\bibitem{Beisert:2010jr}
N.~Beisert, C.~Ahn, L.~F.~Alday, Z.~Bajnok, J.~M.~Drummond et~al.,
\textit{``{Review of AdS/CFT Integrability: An Overview}''},
\textsf{\doiref{10.1007/s11005-011-0529-2}{Lett.Math.Phys.~99,~3~(2012)}},
\texttt{\arxivref{1012.3982}{arxiv:1012.3982}}.

\bibitem{Escobedo:2010xs}
J.~Escobedo, N.~Gromov, A.~Sever and P.~Vieira,
\textit{``{Tailoring Three-Point Functions and Integrability}''},
\textsf{\doiref{10.1007/JHEP09(2011)028}{JHEP~1109,~028~(2011)}},
\texttt{\arxivref{1012.2475}{arxiv:1012.2475}}.

\bibitem{Escobedo:2011xw}
J.~Escobedo, N.~Gromov, A.~Sever and P.~Vieira,
\textit{``{Tailoring Three-Point Functions and Integrability II. Weak/strong
  coupling match}''},
\textsf{\doiref{10.1007/JHEP09(2011)029}{JHEP~1109,~029~(2011)}},
\texttt{\arxivref{1104.5501}{arxiv:1104.5501}}.

\bibitem{Gromov:2011jh}
N.~Gromov, A.~Sever and P.~Vieira,
\textit{``{Tailoring Three-Point Functions and Integrability III. Classical
  Tunneling}''},
\textsf{\doiref{10.1007/JHEP07(2012)044}{JHEP~1207,~044~(2012)}},
\texttt{\arxivref{1111.2349}{arxiv:1111.2349}}.

\bibitem{Foda:2011rr}
O.~Foda,
\textit{``{N=4 SYM structure constants as determinants}''},
\textsf{\doiref{10.1007/JHEP03(2012)096}{JHEP~1203,~096~(2012)}},
\texttt{\arxivref{1111.4663}{arxiv:1111.4663}}.

\bibitem{Kazakov:2012ar}
V.~Kazakov and E.~Sobko,
\textit{``{Three-point correlators of twist-2 operators in N=4 SYM at Born
  approximation}''},
\textsf{\doiref{10.1007/JHEP06(2013)061}{JHEP~1306,~061~(2013)}},
\texttt{\arxivref{1212.6563}{arxiv:1212.6563}}.

\bibitem{Foda:2013nua}
O.~Foda, Y.~Jiang, I.~Kostov and D.~Serban,
\textit{``{A tree-level 3-point function in the su(3)-sector of planar N=4
  SYM}''},
\texttt{\arxivref{1302.3539}{arxiv:1302.3539}}.

\bibitem{Drummond:2010km}
J.~Drummond,
\textit{``{Review of AdS/CFT Integrability, Chapter V.2: Dual Superconformal
  Symmetry}''},
\textsf{\doiref{10.1007/s11005-011-0519-4}{Lett.Math.Phys.~99,~481~(2012)}},
\texttt{\arxivref{1012.4002}{arxiv:1012.4002}}.

\bibitem{Drummond:2008vq}
J.~M.~Drummond, J.~Henn, G.~P.~Korchemsky and E.~Sokatchev,
\textit{``{Dual superconformal symmetry of scattering amplitudes in
  {$\mathcal{N}=\mathord{}$4} super-Yang--Mills theory}''},
\textsf{\doiref{10.1016/j.nuclphysb.2009.11.022}{Nucl.~Phys.~B828,~317~(2010)}},
\texttt{\arxivref{0807.1095}{arxiv:0807.1095}}.

\bibitem{Drummond:2009fd}
J.~M.~Drummond, J.~M.~Henn and J.~Plefka,
\textit{``{Yangian symmetry of scattering amplitudes in
  {$\mathcal{N}=\mathord{}$4} super Yang-Mills theory}''},
\textsf{\doiref{10.1088/1126-6708/2009/05/046}{JHEP~0905,~046~(2009)}},
\texttt{\arxivref{0902.2987}{arxiv:0902.2987}}.

\bibitem{ArkaniHamed:2010kv}
N.~Arkani-Hamed, J.~L.~Bourjaily, F.~Cachazo, S.~Caron-Huot and J.~Trnka,
\textit{``{The All-Loop Integrand For Scattering Amplitudes in Planar N=4
  SYM}''},
\textsf{\doiref{10.1007/JHEP01(2011)041}{JHEP~1101,~041~(2011)}},
\texttt{\arxivref{1008.2958}{arxiv:1008.2958}}.

\bibitem{Beisert:2010gn}
N.~Beisert, J.~Henn, T.~McLoughlin and J.~Plefka,
\textit{``{One-Loop Superconformal and Yangian Symmetries of Scattering
  Amplitudes in N=4 Super Yang-Mills}''},
\textsf{\doiref{10.1007/JHEP04(2010)085}{JHEP~1004,~085~(2010)}},
\texttt{\arxivref{1002.1733}{arxiv:1002.1733}}.

\bibitem{Sever:2009aa}
A.~Sever and P.~Vieira,
\textit{``{Symmetries of the N=4 SYM S-matrix}''},
\texttt{\arxivref{0908.2437}{arxiv:0908.2437}}.

\bibitem{Alday:2007hr}
L.~F.~Alday and J.~M.~Maldacena,
\textit{``{Gluon scattering amplitudes at strong coupling}''},
\textsf{\doiref{10.1088/1126-6708/2007/06/064}{JHEP~0706,~064~(2007)}},
\texttt{\arxivref{0705.0303}{arxiv:0705.0303}}.

\bibitem{Brandhuber:2007yx}
A.~Brandhuber, P.~Heslop and G.~Travaglini,
\textit{``{MHV Amplitudes in {$\mathcal{N}=\mathord{}$4} Super Yang--Mills and
  Wilson Loops}''},
\textsf{\doiref{10.1016/j.nuclphysb.2007.11.002}{Nucl.~Phys.~B794,~231~(2008)}},
\texttt{\arxivref{0707.1153}{arxiv:0707.1153}}.

\bibitem{Drummond:2007cf}
J.~M.~Drummond, J.~Henn, G.~P.~Korchemsky and E.~Sokatchev,
\textit{``{On planar gluon amplitudes/Wilson loops duality}''},
\textsf{\doiref{10.1016/j.nuclphysb.2007.11.007}{Nucl.~Phys.~B795,~52~(2008)}},
\texttt{\arxivref{0709.2368}{arxiv:0709.2368}}.

\bibitem{CaronHuot:2010ek}
S.~Caron-Huot,
\textit{``{Notes on the scattering amplitude / Wilson loop duality}''},
\textsf{\doiref{10.1007/JHEP07(2011)058}{JHEP~1107,~058~(2011)}},
\texttt{\arxivref{1010.1167}{arxiv:1010.1167}}.

\bibitem{Mason:2010yk}
L.~Mason and D.~Skinner,
\textit{``{The Complete Planar S-matrix of N=4 SYM as a Wilson Loop in Twistor
  Space}''},
\textsf{\doiref{10.1007/JHEP12(2010)018}{JHEP~1012,~018~(2010)}},
\texttt{\arxivref{1009.2225}{arxiv:1009.2225}}.

\bibitem{Belitsky:2012nu}
A.~Belitsky,
\textit{``{Conformal anomaly of super Wilson loop}''},
\textsf{\doiref{10.1016/j.nuclphysb.2012.04.022}{Nucl.Phys.~B862,~430~(2012)}},
\texttt{\arxivref{1201.6073}{arxiv:1201.6073}}.

\bibitem{CaronHuot:2011ky}
S.~Caron-Huot,
\textit{``{Superconformal symmetry and two-loop amplitudes in planar N=4 super
  Yang-Mills}''},
\textsf{\doiref{10.1007/JHEP12(2011)066}{JHEP~1112,~066~(2011)}},
\texttt{\arxivref{1105.5606}{arxiv:1105.5606}}.

\bibitem{Beisert:2012xx}
N.~Beisert, S.~He, B.~U.~Schwab and C.~Vergu,
\textit{``{Null Polygonal Wilson Loops in Full N=4 Superspace}''},
\textsf{\doiref{10.1088/1751-8113/45/26/265402}{J.Phys.~A45,~265402~(2012)}},
\texttt{\arxivref{1203.1443}{arxiv:1203.1443}}.

\bibitem{Basso:2013vsa}
B.~Basso, A.~Sever and P.~Vieira,
\textit{``{Space-time S-matrix and Flux-tube S-matrix at Finite Coupling}''},
\texttt{\arxivref{1303.1396}{arxiv:1303.1396}}.

\bibitem{Basso:2013aha}
B.~Basso, A.~Sever and P.~Vieira,
\textit{``{Space-time S-matrix and Flux tube S-matrix II. Extracting and
  Matching Data}''},
\texttt{\arxivref{1306.2058}{arxiv:1306.2058}}.

\bibitem{Maldacena:1997re}
J.~M.~Maldacena,
\textit{``{The Large N limit of superconformal field theories and
  supergravity}''},
\textsf{Adv.Theor.Math.Phys.~2,~231~(1998)},
\texttt{\arxivref{hep-th/9711200}{hep-th/9711200}}.

\bibitem{Maldacena:1998im}
J.~M.~Maldacena,
\textit{``{Wilson loops in large N field theories}''},
\textsf{\doiref{10.1103/PhysRevLett.80.4859}{Phys.Rev.Lett.~80,~4859~(1998)}},
\texttt{\arxivref{hep-th/9803002}{hep-th/9803002}}.

\bibitem{Rey:1998ik}
S.-J.~Rey and J.-T.~Yee,
\textit{``{Macroscopic strings as heavy quarks in large N gauge theory and
  anti-de Sitter supergravity}''},
\textsf{\doiref{10.1007/s100520100799}{Eur.Phys.J.~C22,~379~(2001)}},
\texttt{\arxivref{hep-th/9803001}{hep-th/9803001}}.

\bibitem{Polyakov:2000ti}
A.~M.~Polyakov and V.~S.~Rychkov,
\textit{``{Gauge field strings duality and the loop equation}''},
\textsf{\doiref{10.1016/S0550-3213(00)00183-8}{Nucl.Phys.~B581,~116~(2000)}},
\texttt{\arxivref{hep-th/0002106}{hep-th/0002106}}.

\bibitem{Polyakov:2000jg}
A.~M.~Polyakov and V.~S.~Rychkov,
\textit{``{Loop dynamics and AdS / CFT correspondence}''},
\textsf{\doiref{10.1016/S0550-3213(00)00642-8}{Nucl.Phys.~B594,~272~(2001)}},
\texttt{\arxivref{hep-th/0005173}{hep-th/0005173}}.

\bibitem{Polyakov-unpublished}
A.~M.~Polyakov,
\textit{``{unpublished}''}.

\bibitem{Erickson:2000af}
J.~K.~Erickson, G.~W.~Semenoff and K.~Zarembo,
\textit{``Wilson loops in {$\mathcal{N}=\mathord{}$4} supersymmetric Yang-Mills
  theory''},
\textsf{\doiref{10.1016/S0550-3213(00)00300-X}{Nucl.~Phys.~B582,~155~(2000)}},
\texttt{\arxivref{hep-th/0003055}{hep-th/0003055}}.

\bibitem{Drukker:2000rr}
N.~Drukker and D.~J.~Gross,
\textit{``An exact prediction of {$\mathcal{N}=\mathord{}$4} SUSYM theory for
  string theory''},
\textsf{J.~Math.~Phys.~42,~2896~(2001)},
\texttt{\arxivref{hep-th/0010274}{hep-th/0010274}}.

\bibitem{Polyakov:1979gp}
A.~M.~Polyakov,
\textit{``{String Representations and Hidden Symmetries for Gauge Fields}''},
\textsf{\doiref{10.1016/0370-2693(79)90747-0}{Phys.Lett.~B82,~247~(1979)}}.

\bibitem{Polyakov:1980ca}
A.~M.~Polyakov,
\textit{``{Gauge Fields as Rings of Glue}''},
\textsf{\doiref{10.1016/0550-3213(80)90507-6}{Nucl.~Phys.~B164,~171~(1980)}}.

\bibitem{Makeenko:1979pb}
Y.~Makeenko and A.~A.~Migdal,
\textit{``{Exact Equation for the Loop Average in Multicolor QCD}''},
\textsf{\doiref{10.1016/0370-2693(79)90131-X}{Phys.Lett.~B88,~135~(1979)}}.

\bibitem{Makeenko:1980vm}
Y.~Makeenko and A.~A.~Migdal,
\textit{``{Quantum Chromodynamics as Dynamics of Loops}''},
\textsf{\doiref{10.1016/0550-3213(81)90258-3}{Nucl.Phys.~B188,~269~(1981)}}.

\bibitem{Drukker:1999zq}
N.~Drukker, D.~J.~Gross and H.~Ooguri,
\textit{``{Wilson loops and minimal surfaces}''},
\textsf{\doiref{10.1103/PhysRevD.60.125006}{Phys.~Rev.~D60,~125006~(1999)}},
\texttt{\arxivref{hep-th/9904191}{hep-th/9904191}}.

\bibitem{Drukker:1999gy}
N.~Drukker,
\textit{``{A new type of loop equations}''},
\textsf{\doiref{10.1088/1126-6708/1999/11/006}{JHEP~9911,~006~(1999)}},
\texttt{\arxivref{hep-th/9908113}{hep-th/9908113}}.

\bibitem{MacKay:2004tc}
N.~MacKay,
\textit{``{Introduction to Yangian symmetry in integrable field theory}''},
\textsf{\doiref{10.1142/S0217751X05022317}{Int.J.Mod.Phys.~A20,~7189~(2005)}},
\texttt{\arxivref{hep-th/0409183}{hep-th/0409183}}.

\bibitem{Drinfeld:1985rx}
V.~G.~Drinfel'd,
\textit{``Hopf algebras and the quantum Yang-Baxter equation''},
\textsf{Sov.~Math.~Dokl.~32,~254~(1985)}.

\bibitem{Drinfeld:1986in}
V.~G.~Drinfel'd,
\textit{``Quantum groups''},
\textsf{\doiref{10.1007/BF01247086}{J.~Math.~Sci.~41,~898~(1988)}}.

\bibitem{Dolan:2004ps}
L.~Dolan, C.~R.~Nappi and E.~Witten,
\textit{``Yangian symmetry in $D=$4 superconformal Yang-Mills theory''},
\texttt{\arxivref{hep-th/0401243}{hep-th/0401243}},
in: \textit{``Quantum Theory and Symmetries''},
ed.: P.~C.~Argyres et~al.,
World Scientific (2004),
Singapore.

\bibitem{Harnad:1985bc}
J.~P.~Harnad and S.~Shnider,
\textit{``{Constraints and field equations for ten-dimensional super Yang-Mills
  theory}''},
\textsf{\doiref{10.1007/BF01454971}{Commun.Math.Phys.~106,~183~(1986)}}.

\bibitem{Semenoff:2002kk}
G.~W.~Semenoff and K.~Zarembo,
\textit{``{Wilson loops in SYM theory: From weak to strong coupling}''},
\textsf{\doiref{10.1016/S0920-5632(02)01312-9}{Nucl.Phys.Proc.Suppl.~108,~106~(2002)}},
\texttt{\arxivref{hep-th/0202156}{hep-th/0202156}}.

\bibitem{Ferro:2012xw}
L.~Ferro, T.~Lukowski, C.~Meneghelli, J.~Plefka and M.~Staudacher,
\textit{``{Harmonic R-matrices for Scattering Amplitudes and Spectral
  Regularization}''},
\textsf{\doiref{10.1103/PhysRevLett.110.121602}{Phys.Rev.Lett.~110,~121602~(2013)}},
\texttt{\arxivref{1212.0850}{arxiv:1212.0850}}.

\bibitem{Ferro:2013dga}
L.~Ferro, T.~Lukowski, C.~Meneghelli, J.~Plefka and M.~Staudacher,
\textit{``{Spectral Parameters for Scattering Amplitudes in N=4 Super
  Yang-Mills Theory}''},
\texttt{\arxivref{1308.3494}{arxiv:1308.3494}}.

\bibitem{ArkaniHamed:2008gz}
N.~Arkani-Hamed, F.~Cachazo and J.~Kaplan,
\textit{``{What is the Simplest Quantum Field Theory?}''},
\texttt{\arxivref{0808.1446}{arxiv:0808.1446}}.

\bibitem{Brandhuber:2008pf}
A.~Brandhuber, P.~Heslop and G.~Travaglini,
\textit{``{A note on dual superconformal symmetry of the
  {$\mathcal{N}=\mathord{}$4} super Yang-Mills S-matrix}''},
\textsf{\doiref{10.1103/PhysRevD.78.125005}{Phys.~Rev.~D78,~125005~(2008)}},
\texttt{\arxivref{0807.4097}{arxiv:0807.4097}}.

\bibitem{Elvang:2008na}
H.~Elvang, D.~Z.~Freedman and M.~Kiermaier,
\textit{``{Recursion Relations, Generating Functions, and Unitarity Sums in
  {$\mathcal{N}=\mathord{}$4} SYM Theory}''},
\textsf{\doiref{10.1088/1126-6708/2009/04/009}{JHEP~0904,~009~(2009)}},
\texttt{\arxivref{0808.1720}{arxiv:0808.1720}}.

\bibitem{Drummond:2008cr}
J.~M.~Drummond and J.~M.~Henn,
\textit{``{All tree-level amplitudes in {$\mathcal{N}=\mathord{}$4} SYM}''},
\textsf{\doiref{10.1088/1126-6708/2009/04/018}{JHEP~0904,~018~(2009)}},
\texttt{\arxivref{0808.2475}{arxiv:0808.2475}}.

\bibitem{Ishizeki:2011bf}
R.~Ishizeki, M.~Kruczenski and S.~Ziama,
\textit{``{Notes on Euclidean Wilson loops and Riemann Theta functions}''},
\textsf{\doiref{10.1103/PhysRevD.85.106004}{Phys.Rev.~D85,~106004~(2012)}},
\texttt{\arxivref{1104.3567}{arxiv:1104.3567}}.

\bibitem{Alday:2010vh}
L.~F.~Alday, J.~Maldacena, A.~Sever and P.~Vieira,
\textit{``{Y-system for Scattering Amplitudes}''},
\textsf{\doiref{10.1088/1751-8113/43/48/485401}{J.Phys.~A43,~485401~(2010)}},
\texttt{\arxivref{1002.2459}{arxiv:1002.2459}}.

\bibitem{Belitsky:2003sh}
A.~V.~Belitsky, S.~E.~Derkachov, G.~Korchemsky and A.~Manashov,
\textit{``{Superconformal operators in N=4 superYang-Mills theory}''},
\textsf{\doiref{10.1103/PhysRevD.70.045021}{Phys.Rev.~D70,~045021~(2004)}},
\texttt{\arxivref{hep-th/0311104}{hep-th/0311104}}.

\bibitem{Howe:1996rk}
P.~S.~Howe and P.~C.~West,
\textit{``{Superconformal invariants and extended supersymmetry}''},
\textsf{\doiref{10.1016/S0370-2693(97)00340-7}{Phys.Lett.~B400,~307~(1997)}},
\texttt{\arxivref{hep-th/9611075}{hep-th/9611075}}.

\end{thebibliography}

\end{document}